\documentclass[twocolumn]{svjour3}

\usepackage{subcaption}
\usepackage{SpringerLarge}
\usepackage{orcidlink}
\usepackage[misc]{ifsym} 
\usepackage{xspace}
\usepackage{listings}
\usepackage{lstautogobble}
\usepackage{amsmath}
\usepackage{cleveref}
\usepackage{float}
\usepackage{fancyvrb}

\newcommand\Real{\mathbb{R}}
\newcommand\Bool{\mathbb{B}}
\newcommand{\expr}{\textit{expr}}
\newcommand{\is}{i}
\newcommand{\numIs}{I}
\newcommand{\os}{o}
\newcommand{\numOs}{O}
\newcommand{\cs}{c}
\newcommand{\numCs}{C}
\newcommand{\trs}{t}
\newcommand{\numTs}{T}
\newcommand{\csv}{\cs^V}
\newcommand{\numCsv}{{V_c}}
\newcommand{\osv}{\os^V}
\newcommand{\numOsv}{{V_o}}

\definecolor{errorred}{HTML}{ab0303}

\definecolor{bluekeywords}{rgb}{0.13, 0.13, 1}
\definecolor{greentypes}{rgb}{0, 0.5, 0}
\definecolor{inferedgreentypes}{rgb}{1.0, 0.2, 0}
\definecolor{orangecomments}{rgb}{1, 0.5, 0.1}
\definecolor{redstrings}{RGB}{171, 114, 2}
\definecolor{graynumbers}{rgb}{0.5, 0.5, 0.5}
\definecolor{goldcomments}{rgb}{0.6, 0.4, 0.08}

\lstdefinelanguage{Lola}{
  keywords=[0]{input, output, trigger, constant, import, spawn, eval, close, with, when},
  moredelim=**[is][\transparent{0.6}]{?}{?},
  moredelim=**[is][\color{greentypes}@]{@}{@},
  keywordstyle=[0]\bfseries\color{bluekeywords},
  keywords=[1]{if, then, else, aggregate, defaults, offset, last, by, or, to, sin, cos, abs, hold, over, using, over_instances},
  keywords=[2]{Variable, String, Int, Int64, UInt, UInt64, Bool, Float32, Float64, Float},
  keywordstyle=[2]\color{greentypes},
  sensitive=false,
  comment=[l]{//},
  morecomment=[s]{/*}{*/},
  morestring=[b]',
  morestring=[b]",
  literate={\\@}{@}1
}
\lstdefinelanguage{SMT}{
  keywords=[0]{assert, declare-fun, check-sat, ion},
  moredelim=**[is][\transparent{0.6}]{?}{?},
  moredelim=**[is][\color{greentypes}@]{@}{@},
  keywordstyle=[0]\bfseries\color{bluekeywords},
  keywords=[2]{Int, Real},
  keywordstyle=[2]\color{greentypes},
  escapeinside={(*@}{@*)},
  sensitive=false,
  comment=[l]{;},
  morestring=[b]',
  morestring=[b]",
  literate={\\@}{@}1,
  alsoletter=-_/,
}

\journalname{}
\title{Stream-based Online and Offline Monitoring under Measurement Noise\thanks{This work was partially supported by the German Research Foundation (DFG) as part of PreCePT (FI 936/7-1; FR 2715/6-1) and of TRR 248 (No. 389792660), by the State of Lower Saxony within the Zukunftslabor Mobilität, and by the European Research Council (ERC) Grant HYPER (No. 101055412).}}

\date{}
\doi{}

\author{Bernd Finkbeiner\orcidlink{0000-0002-4280-8441} \and
Martin Fränzle\orcidlink{0000-0002-9138-8340} \and
Florian Kohn\orcidlink{0000-0001-9672-2398} \and
Paul Kröger\orcidlink{0000-0002-0301-3611}}

\institute{
Bernd Finkbeiner \at
CISPA Helmholtz Center for Information Security, Germany\\
Technical University of Munich, Germany\\
\email{finkbeiner@cispa.de}
\and
Florian Kohn (\Letter) \at
CISPA Helmholtz Center for Information Security, Germany\\
\email{florian.kohn@cispa.de}
\and
Martin Fränzle \and Paul Kröger \at
Carl von Ossietzky Universität, Oldenburg, Germany\\
\email{\{martin.fraenzle, paul.kroeger\}@uol.de}
}
\authorrunning{Finkbeiner et al.}

\newcommand{\rlola}{\texttt{RLola}\xspace}

\begin{document}
\abstract{
Stream-based monitoring is a runtime verification approach for cyber-physical systems that translates streams of input data, such as sensor readings, into streams of aggregate statistics and verdicts about the safety of the system. 
It is usually assumed that the values on the input streams represent fully accurate measurements of the
physical world. In reality, however, physical sensors are prone to measurement noise and errors.
These errors are further amplified by the processing and aggregation steps within the monitor.
This paper introduces \rlola, a robust extension of the stream-based specification language Lola.
\rlola incorporates the concept of slack variables, which symbolically represent measurement noise while avoiding the aliasing problem of interval arithmetic.

We present algorithms for both online and offline monitoring of \rlola specifications.
Since monitoring \rlola specifications may require unbounded memory in general, we identify a rich fragment of \rlola that can be automatically translated into monitors with guaranteed constant memory usage for online monitoring.
An online \rlola monitor observes a live system and provides real-time feedback on the current status of specified assertions.
A satisfiability-modulo-theories-based offline algorithm analyzes complete system traces and determines whether a hypothetical ground-truth trace exists that satisfies all assertions at all time points.
The offline algorithm can therefore detect violations that the online algorithm may miss.
We implement these algorithms in the existing RTLola framework and evaluate their precision and running time based on a comprehensive example.

}

\maketitle

\section{Introduction} \label{sec:intro}
\begin{figure}
    \centering
    \includegraphics[width=\linewidth]{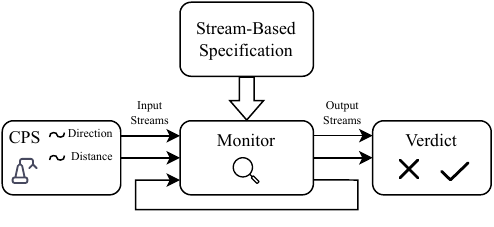}
    \caption{The Stream-Based Monitoring Approach.}
    \label{fig:stream-based}
\end{figure}

Cyber-physical systems (CPS), like autonomous vehicles, drones, power plants, and medical devices, interact with their physical environment through sensors and actuators.
Their operating in safety-critical domains makes assuring their safe behavior safety-critical.
Stream-based monitoring, as visualized in \Cref{fig:stream-based}, is an established runtime verification approach for cyber-physical systems.
Input streams capturing sensor readings are translated into output streams that process and aggregate these measurements.
The resulting values on the output streams are then continuously evaluated against trigger conditions that characterize erroneous or dangerous situations.
Because stream-based monitoring tools such as RTLola \cite{RTLola}, TeSSLa \cite{TeSSLa}, or Striver \cite{Striver} are used in safety-critical applications, such as autonomous aircraft~\cite{cav4}, the precision of the monitoring result is vital.
But how precise are stream-based monitors?

Most frameworks for stream-based monitoring assume that input streams capture fully accurate measurements of the physical reality.
Consequently, output streams are also assumed to be precise, even if they are based on input data collected from physical sensors, which are prone to measurement noise and errors, and even if this data has been processed in a way that may have significantly amplified these errors.

A well-known idea to keep track of measurement noise is to lift individual values from scalars to intervals, akin to interval-based robust monitoring of Signal Temporal Logic specifications \cite{ViscontiEA:IntervalBasedSTRELMonitoring}.
Input streams then produce intervals centered around the measured value with a width defined by the sensor's precision.
The error in the output streams can be tracked via interval arithmetic: as more errors are combined into individual output values, the intervals of these outputs becomes larger, and we can determine whether a trigger condition is violated, given the precision of the data.

Unfortunately, interval analysis usually leads to an overly pessimistic result. This phenomenon is known as the \emph{aliasing} or \emph{dependency problem}: in situations where errors cancel each other out, for example, because the same input is added and later subtracted from an aggregate value, interval arithmetic will still add, rather than subtract, the errors. In the trivial example, the term $x-x$ should evaluate to 0, independently of the value of $x$. However, if, because of noise, we assume $x$ to be in the interval $[{-}10, 10]$, then interval arithmetic produces the even larger interval 
$[{-}10, 10] - [{-}10, 10] = [{-20}, 20]$.

In this paper, we present \rlola (RobustLola), an extension of the Lola monitoring language~\cite{Lola} that explicitly tracks the precision of stream values.
To avoid the aliasing problem, we introduce explicit \emph{slack variables} that represent measurement noise by symbolically representing the interval $[-1, 1]$, which can be extended to arbitrary intervals through affine arithmetic~\cite{AffineArithmetic}.
%Because each variable identifies a particular source of noise, they are not susceptible to the aliasing problem.
In the example, the uncertain input is represented as $x + 10 \epsilon$, where $\epsilon$ is a slack variable.
Subtracting the input from itself would then result in $(x + 10 \epsilon) - (x + 10 \epsilon)=0$.
The advantage of slack variables is, thus, their higher accuracy; a potentially big disadvantage, on the other hand, is an increase in computational cost, such as memory usage. 

In \emph{online} monitoring, the typical requirement is that memory usage remains constant with respect to the length of the input streams, ensuring that the monitor can operate without exceeding its memory capacity.
For \rlola, if the noise in individual input stream values is at least partially independent, a separate slack variable is required for each time point.
Since slack variables are unlikely to reduce to scalar values, the number of slack variables in the monitor’s equation store can grow without bound.

However, for a broad class of practically relevant \rlola specifications, it is possible to combine multiple slack variables into a single variable without loss of precision.
For example, the term $x + 5\epsilon_1 + 5\epsilon_2$ is equivalent to $x+ 10 \epsilon'$ if $\epsilon_1$ and $\epsilon_2$ occurs only there.
Suppose now that $x$ is an output stream in which the term $5 \epsilon$, with fresh slack variable $\epsilon$, is added in every step.
Then, instead of keeping a growing term $5\epsilon_1 + 5\epsilon_2 + \ldots + 5\epsilon_n$ in memory, it suffices to count the number of steps $n$, and replace the $n$ slack variables with a single variable with factor $5n$, resulting in the term $5n \epsilon'$.

Based on these observations, we identify a syntactic fragment of \rlola that can be automatically translated into an equivalent Lola specification.
The resulting monitor has provably constant memory consumption.

This approach, however, entails some loss of accuracy.
Dependent noise, such as a fixed calibration error of a sensor, introduces correlations across different time points.
As a result, a trigger condition that was satisfied in the past may later be falsified.
Accurately capturing such dependencies requires storing the monitoring equations for all previous time points, enabling reevaluation at any later time.
For online monitoring, we accept this loss of accuracy, prioritizing the bounded-memory requirement over retrospective precision.

In \emph{offline} monitoring, by contrast, the entire system trace is available in advance, so only a bounded amount of data must be analyzed.
Without the constant-memory restriction, all \rlola specifications can be evaluated with full accuracy.
We introduce an offline monitoring algorithm for \rlola based on satisfiability modulo theories (SMT), which produces \emph{global} verdicts for fixed traces of arbitrary length.

Semantically, online monitoring evaluates boolean trigger conditions \emph{locally}, verifying \emph{at each time step} that a hypothetical ground-truth trace exists that is consistent with the measurements and avoids the trigger condition.
Offline monitoring, in contrast, analyzes complete system traces, verifying the existence of a hypothetical ground-truth trace that is consistent with the measurements and avoids the trigger conditions \emph{across all time steps}.

In the remainder of this section, we present a motivating example and discuss related work. 
In \Cref{sec:prelims} we review the syntax and semantics of Lola\cite{Lola} before presenting the syntax and semantics of its extension \rlola in \Cref{sec:lang}. 
The online evaluation methods are discussed in \Cref{sec:online}
The SMT-based offline algorithm is introduced in \Cref{sec:offline}.
Before evaluating the algorithms in \Cref{sec:eval}, we discuss the implementation of the algorithms in \Cref{sec:tool}. We conclude with a brief outlook in \Cref{sec:conclusion}.

This paper is an extended version of the conference publication~\cite{DBLP:conf/rv/FinkbeinerFKK24}.
Relative to the conference version, it introduces an SMT-based offline monitoring approach and provides simplified explanations to improve accessibility.
In addition, we provide an implementation of the proposed algorithms within the existing RTLola toolchain~\cite{cav4,DBLP:conf/fm/BaumeisterFKS24}.

\subsection{Motivating Example}\label{sec:motivating}
\begin{figure}
    \centering
    \includegraphics[width=\linewidth]{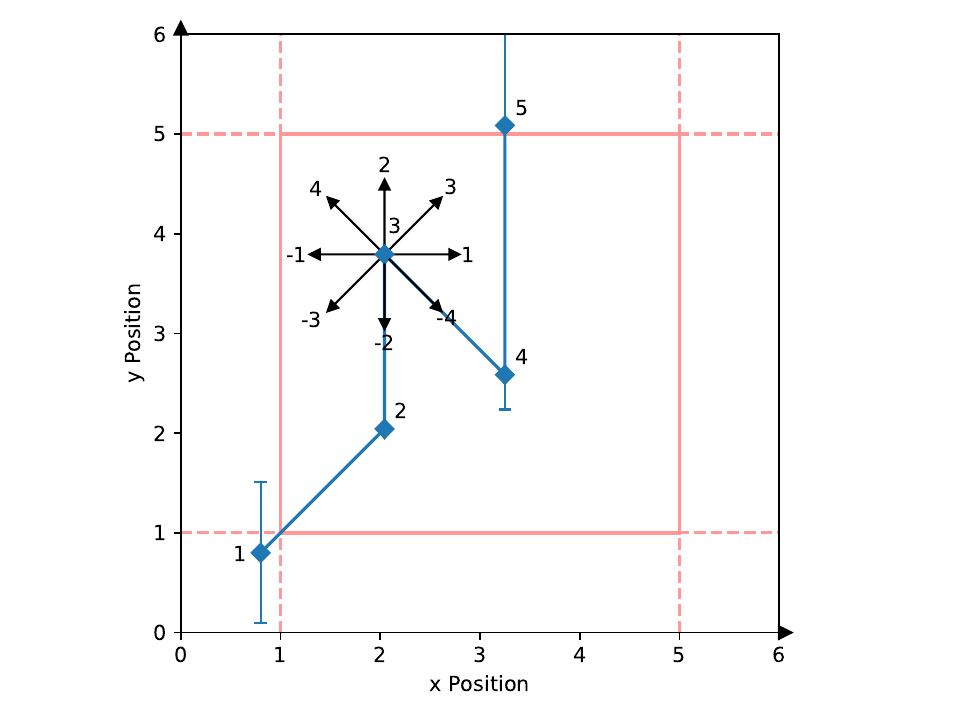}
    \caption{An Example Path of the Robot.}
    \label{fig:running_example}
\end{figure}

As a motivating example, consider an industrial warehouse robot that autonomously navigates by tracking its position through an unreliable indoor positioning system.
The example illustrates the use of slack variables in handling measurement noise and demonstrates how the offline analysis compares to online methods
The industrial robot is capable of moving in four directions, as shown by the arrows in \Cref{fig:running_example}.
The robot tracks its position relative to a fixed starting point $(0,0)$, recording movements as a combination of direction and distance.
The robot moves on a rail system, ensuring precise direction measurements, while distance measurements are subject to calibration and measurement errors.
An \rlola specification capturing this behavior is given in \Cref{ex:motivating} below.
Note that for simplicity, the computation of the $y$ position is omitted as it follows analogously.
\begin{example}[Warehouse Robot]
\label{ex:motivating}
\begin{lstlisting}
input direction: Int
input raw_distance: Float
constant cos_45: Float := (1/2) * sqrt(2)

constant delta: Variable
output e: Variable
output distance := raw_distance + 0.1 * e + 0.9 * delta

output position_x := 
     if direction = 1 then 
        position_x.offset(by: -1, or: 0) + distance
     else if direction = -1 then 
        position_x.offset(by: -1, or: 0) - distance
     else if direction = 3 then 
        position_x.offset(by: -1, or: 0) + cos_45 * distance
     else if direction = -3 then 
        position_x.offset(by: -1, or: 0) - cos_45 * distance
     else if direction = 4 then 
        position_x.offset(by: -1, or: 0) - cos_45 * distance
     else if direction = -4 then 
        position_x.offset(by: -1, or: 0) + cos_45 * distance
     else
        position_x.offset(by: -1, or: 0)

trigger 1.0 $>_{0.65}$ position_x || 5.0 $<_{0.65}$ position_x
\end{lstlisting}
\end{example}
Intuitively, slack variables extend the value domain of streams such that a stream of type \lstinline{Variable} produces a fresh slack variable every time it is computed.
As the precise value of the slack variables cannot be computed, such streams do not require a defining stream equation.
Similarly, a constant of type \lstinline{Variable} represents a single slack variable.
In the above example, we define a stream of slack variables \lstinline{e} and a single slack variable \lstinline{delta} that are added to the \lstinline{raw_distance} in the \lstinline{distance} output stream to capture the measurement noise of the distance measurements.
A more detailed description of this measurement error model is given in \Cref{sec:error-model}.
Depending on the movement direction, this corrected distance is added proportionally to the computed position.

The trigger definitions at the end of the specification determine when the robot exits its safe working area based on its position.
Due to the slack variables, the position streams represent ranges of possible positions.
To avoid inconclusive verdicts, the $>_{0.65}$ and $<_{0.65}$ operators are annotated with a factor, indicating the required overlap of the position range with the threshold for the expression to be true.

\paragraph{Online Monitoring.}
In the above example, all slack variables produced by the \lstinline{e} stream will accumulate in the computed position streams, as the \lstinline{offset(by: -1, or: 0)} operator refers to the last stream value (or 0 if that does not exist yet).
As later shown in \Cref{sec:online}, multiple linearly dependent slack variables can be combined without losing precision.
Based on this, building a finite memory online monitor for the above specification is possible.
In fact, it is possible to translate the specification to one without slack variables:
\begin{lstlisting}
input direction: Int
input raw_distance: Float
constant cos_45: Float := (1/2) * sqrt(2)

output d1 := if direction = 1
            then d1.offset(by: -1, or: 0.0) + 1.0
            else d1.offset(by: -1, or: 0.0)
output dn1 := if direction = -1
            then dn1.offset(by: -1, or: 0.0) + 1.0
            else dn1.offset(by: -1, or: 0.0)
output d3, dn3, d4, dn4 := ...

output delta_x := d1 - dn1 + d3*cos_45 - dn3*cos_45 - d4*cos_45 + dn4*cos_45
output epsilon_x := d1 + dn1 + d3 + dn3 + d4 + dn4
output raw_position_x := ...

output position_x_lower := raw_position_x - epsilon_x - delta_x
output position_x_upper := raw_position_x + epsilon_x + delta_x
trigger (1.0 - position_x_lower) / (position_x_upper - position_x_lower) > 0.65
trigger (position_x_upper - 5.0) / (position_x_upper - position_x_lower) > 0.65
\end{lstlisting}
Note that this specification is optimized by removing unreachable or equivalent cases.
The \lstinline{raw_position_x} is defined analogously to the \lstinline{position_x} stream in \Cref{ex:motivating}, with the difference that it references the \lstinline{raw_distance} input stream instead of the \lstinline{distance} stream.
Conceptually, this specification replaces the stream of slack variables \lstinline{e} with eight individual slack variables, one for each direction.
The streams \lstinline{d1} to \lstinline{d4} and their negative counterparts dynamically compute the coefficients of these slack variables, while the \lstinline{delta_x} stream tracks the coefficient of the \lstinline{delta} slack variable.
The streams \lstinline{position_x_lower} and \lstinline{position_x_upper} reconstruct a precise lower and upper bound of the original \lstinline{position_x} stream by applying interval arithmetic.
In \Cref{sec:online}, we define a syntactical fragment of \rlola for which such a transformation is always possible. 

\paragraph{Offline Monitoring.} Offline monitoring is characterized by a finite trace and virtually infinite computational resources, allowing for more precise disambiguation of trigger conditions than the overlap percentages used in online monitoring.
Instead of using overlap percentages, we propose evaluating all trigger conditions either existentially or universally over all time steps for offline monitoring.

\rlola specifications essentially reason about a set of hypothetical ground truth traces compatible with runtime measurements. Compatibility, defined in \Cref{sec:error-model}, implies the existence of a slack variable assignment that equates measurements to the ground truth trace.
Existential evaluation of trigger conditions means that at least one compatible ground truth trace must not violate any trigger.

The difference in trigger evaluation between online and offline monitoring leads to more precise verdicts from the offline algorithm, as illustrated in \Cref{fig:running_example}.
This figure plots an execution trace of the monitor, showing the robot's computed x and y positions.
Error bars at points one and five indicate the y-position error margins calculated by the online monitoring algorithm.
The online algorithm deems the specification satisfied because less than 65\% of the error margins at points one and five overlap the threshold at each individual time step.
However, the offline analysis, considering trigger conditions across all time points, proves the specification unsatisfiable.
This is because the measurement at point one constrains the shared calibration error \lstinline{delta} to be positive, which contradicts the measurement at point five.
In \Cref{sec:offline}, we present an SMT encoding of \rlola specifications that enables the automatic reasoning about trigger conditions using an SMT solver.

\subsection{Related Work}
There are numerous temporal logics used in runtime monitoring, for example signal temporal logic~\cite{DBLP:conf/formats/MalerN04}, linear temporal logic \cite{DBLP:journals/tosem/BauerLS11}, and metric temporal logic~\cite{DBLP:journals/entcs/ThatiR05}.
Stream-based specification languages like Tessla~\cite{TeSSLa}, Striver~\cite{Striver} or Lola~\cite{Lola,cav4} combine the desirable computational properties of logic with the expressiveness of a programming language, using streams of values as their fundamental computational paradigm.

As established by Kauffman et al.~\cite{DBLP:journals/sttt/KauffmanHF21}, there are properties that cannot be monitored over unreliable channels that alter, delay, or lose data.
While previous work was focused on missing or shifted events \cite{DBLP:conf/rv/LeuckerSS0T19}, this paper targets events that are present but mutated by measurement noise as specified by an error model.

While this could be encoded in first-order logic~\cite{DBLP:conf/tacas/DeckerLT14}, the goal is to keep a strict memory bound on the resulting monitor.
Kallwies et al. \cite{DBLP:conf/atva/KallwiesLS22} handle missing events in the stream-based setting through symbolic input variables.
The authors show that, in general, if streams are defined over real and boolean values, precise monitoring requires unbounded memory.
Yet, this paper identifies a syntactic fragment in this domain that still allows for bounded memory monitors.

For temporal logics, there exist robust quantitative interpretations~\cite{DonzeMaler10,DBLP:conf/cav/DonzeFM13,DBLP:conf/ictac/FranzleH05} that can handle inaccurate timestamps or measurement noise. Most such logics do not support measurement-related error models, leading to pessimistic verdicts.
One exception is ``truly robust'' monitoring of Signal Temporal Logic~\cite{RobustSTL}, which also uses slack variables to express error models, but is significantly less expressive than stream-based specification languages.

The precise offline algorithm for \rlola presented in this paper is based on a satisfiability-modulo-theories~\cite{DBLP:reference/mc/BarrettT18} encoding.
Our approach thus extends similar such encodings for temporal logics like STL~\cite{RobustSTL} or LTL~\cite{DBLP:conf/tacas/DeckerLT14} to stream-based monitoring.

\section{Preliminaries}\label{sec:prelims}
Runtime monitoring validates observed system behavior against a formal specification at runtime.
The approach presented in this paper adapts the stream-based specification language Lola \cite{Lola}.
In the following, we provide an overview of its definition.

\subsection{Lola}
\label{sec:Lola}
A Lola~\cite{Lola} specification consists of input streams, representing the observations made of the system, and output streams, which compute new values from input streams and other output streams.
A Lola specification is a set of (recursive) equations over stream variables of the form:
\begin{align*}
o_1 &= \expr_1(i_1,...,i_m,\quad o_1,...,o_n)\\
&\vdots\\
o_n &= \expr_n(i_1, ..., i_m, \quad o_1,...,o_n)\\
\end{align*}
where $o_1,...,o_n$ are output stream variables and $i_1,...,i_m$ are input stream variables and $\expr_1,...,\expr_n$ are stream expressions.

\emph{Stream expressions} determine how the next value of an output stream is computed.
They are defined as arithmetic and logic expressions over stream variables. 
They include \lstinline{if ... then ... else ...} clauses, stream variable references, and a stream offset operator: \lstinline{.offset(by: $i$, or: $l$)} for $i < 0 \in \mathbb{Z}$ and some literal $l$.

Further, we use \lstinline{.last(or: $l$)} as syntactic sugar for \lstinline{.offset(by: -1, or $l$)}.

\emph{The semantics} of Lola is defined by an evaluation model that relates input stream values to output stream values.
For its full definition, we refer to the original Lola paper~\cite{Lola}.
Notice that a specification can have multiple valid evaluation models. For example, the specification: \lstinline{output o = o} has infinitely many evaluation models for any given vector of input streams $\tau_1,\ldots,\tau_n$.
Such specifications are called not well-defined due to their non-determinism.
A specification is well-defined only if it assigns precisely one evaluation model to each vector of input streams $\tau_1,\ldots,\tau_n$.
A syntactic criterion for well-definedness is given with the help of a dependency graph:
\begin{definition}[Dependency Graph]
    Let $\phi$ be a Lola specification. The dependency graph of $\phi$ is a directed weighted multi-graph $G = \langle V, E \rangle$ with \linebreak $V=\{i_1,...,i_m,o_1,...,o_n\}$. An edge $e=\langle o_i, o_k, w \rangle$ is in $E$ iff the expression of $o_i$ contains $o_k.\textit{offset}(\textit{by: }w,\textit{ or: } c)$ as a sub-expression (or $e=\langle o_i, i_k, w \rangle$ if $i_k.\textit{offset}(\textit{by: }w,\textit{ or: } c)$ is a sub-expression). Analogously, edges with weight 0 are added for non-offset accesses.
\end{definition}
A specification is labeled \emph{well-formed} iff its dependency graph does not contain any non-negative weight cycle, where the weight of a cycle is defined as the sum 
of all its edge weights.

\subsection{Error Model}\label{sec:error-model}
Following \cite{RobustSTL}, we adapt the error model induced by the ISO norm 5725 \cite{ISO:5725} by decomposing the measurement error into a constant, but unknown per-sensor offset  and a randomly varying per-measurement error.
This decomposition directly correlates with the ``trueness'' and ``precision'' described in the ISO 5725 standard and is also reflected in \Cref{ex:motivating} through the constant \lstinline{delta} slack variable and the \lstinline{e} stream of fresh slack variables.
We adopt the definition of consistency from \cite{RobustSTL} and define when a series of sensor measurements is consistent with the unknown 
ground truth of a physical property.
\begin{definition}[Consistency]
\label{def:consistency}
Let $S$ be a sensor measuring a physical property at times $T\subseteq\mathbb{N}$ with a maximal sensor offset of $\delta \ge 0$ and a maximal random measurement error of $\epsilon \ge 0$. Let $\tau$ be the ground-truth time series. Then $m_S:T \to \mathbb{R}$ is a possible \emph{$S$ time series over $\tau$} of sensor measurements iff 
\[
 \exists \Delta \in [-\delta,\delta]: \forall t \in T: \exists \varepsilon \in [-\epsilon,\epsilon]: \tau(t) + \varepsilon + \Delta = m_{S}(t).
\]
We say the trajectory \emph{$\tau$ is consistent with $m_S$} and denote this fact by $m_S \models \tau$.
\end{definition}
Note that the consistency relation, as defined above, can be rewritten using affine arithmetic \cite{AffineArithmetic} as: $\tau(t) + \epsilon e_t + \delta d = m_S(t)$ if $d$
is a slack constant and $e_t$ is a per time-step fresh slack variable, ranging over the interval $[-1,1]$.

\section{Robust Lola}\label{sec:lang}
This section defines the syntax and semantics of \rlola (\emph{robust} Lola).
\rlola extends Lola with symbolic slack variables to represent error margins.

To generate slack variables in \rlola, we introduce the \textit{Variable} value type.
An output stream of type \textit{Variable} will produce a new slack variable in each time step or, in case of a constant stream, a single slack variable for the entire execution of the monitor.

Since slack variables are not explicitly bound to any values, streams of type \textit{Variable} have no stream expressions.
Instead, the variables symbolically represent a value in the range $[-1,1]$.
With streams of type \textit{Variable}, we can implement the measurement error model from \Cref{sec:error-model} (like many other error models).

\subsection{\rlola Syntax}
An \rlola specification is defined as a set of (recursive) equations over stream and slack variables as follows:
\begin{definition}[\rlola Specification]\label{def:syntax}
Let $\is_1,...,\is_\numIs$ be input stream variables, $\os_1,...,\os_\numOs$ be output stream variables, $\cs_1, ..., \cs_\numCs$ be constants, $\trs_1,...,\trs_\numTs$ be trigger conditions, $\csv_1,...,\csv_\numCsv$ be constant slack variables, $\osv_1,...,\osv_\numOsv$ be slack variable streams, $\expr_1,...,\expr_\numOs$ be stream expressions, and $C_1,...,C_\numCs \in \Real \cup \Bool$ be constant literals.
\begin{alignat*}{2}
    \cs_1 &:= C_1&&\\
    \vdots & &&\\
    \cs_\numCs &:= C_\numCs\\
    \os_1 &:= \expr_1(\is_1,...,\is_\numIs,\quad \os_1,...,&&\os_\numOs,\quad \cs_1, ...,\cs_\numCs,\\
    & &&\csv_1, ..., \csv_\numCsv, \quad \osv_1, ..., \osv_\numOsv)\\
    \vdots & &&\\
    \os_\numOs &:= \expr_\numOs(\is_1,...,\is_\numIs,\quad \os_1,...,&&\os_\numOs,\quad \cs_1, ..., \cs_\numCs,\\
    & &&\csv_1, ..., \csv_\numCsv, \quad \osv_1, ..., \osv_\numOsv)\\
\end{alignat*}
Omitted from the definition above are \emph{trigger streams}.
They are defined as boolean output streams specifying assertions that are communicated to a system operator upon violation.
\end{definition}
Stream expressions algorithmically define how the next value of an output stream is computed.
They are defined as arithmetic and logical expressions over stream variables, constants, and slack variables as follows:
\begin{definition}[Stream Expression]\label{def:stream_expr}
Let $\expr_1, ..., \expr_k$ be stream expressions of type $T_1, ..., T_k$, then:
\begin{itemize}
    \item If $l$ is a literal of type $T$, then $l$ is a stream expression of type $T$.
    \item If $s$ is a variable of type $T$, then $s$ is a stream expression of type $T$.
    \item If $f: T_1 \times ... \times T_k \rightarrow T$ is a k-ary operator, then $f(\expr_1,...,\expr_k)$ is a stream expression of type $T$.
    \item If $b$ is a boolean stream expression and $\expr_1, \expr_2$ are stream expressions of type $T_1$ and $T_2$ respectively with $T_1 = T_2$, then \lstinline{if } $b$ \lstinline{ then } $\expr_1$ \lstinline{ else } $\expr_2$ is a stream expression of type $T_1$.
    \item If $s$ is an input or output stream variable of type $T$, $i$ is a negative integer and $\expr$ is an expression of type $T$, then $s$\lstinline{.offset}$($\lstinline{by: }$i$,\lstinline{ or: }$\expr)$ is a stream expression of type $T$.
\end{itemize}
\end{definition}
The \emph{dependency graph} and \emph{well-formed} specifications are defined as for Lola \cite{Lola}. 

\subsection{\rlola Semantics}\label{sec:semantics}
As for Lola, the semantics for \rlola is defined by an evaluation model connecting input stream values to output stream values.
Let $\phi$ be a robust Lola specification with:
input stream variables $\is_1,...,\is_\numIs$,
output stream variables $\os_1,...,\os_\numOs$,
constants $\cs_1, ..., \cs_\numCs$,
constant slack variables $\csv_1,...,\csv_\numCsv$,
and slack variable streams $\osv_1,...,\osv_\numOsv$.

We define their semantic counterpart as:
Let $\tau_1,...,\tau_\numIs$ be streams of length $N$ of input values.
Let $\sigma_1,...,\sigma_\numOs$ be streams of length $N$ of output values.
Let $\zeta_1,...,\zeta_\numCsv$ be streams of length $N$ of constant values.
Let $\sigma^V_1,...,\sigma^V_\numOsv$ be streams of length $N$ of slack values.
A single evaluation of $\phi$ is then defined as: $\psi := \{\tau_1, ..., \tau_\numIs,\sigma_1,...,\sigma_\numOs, \sigma^V_1,...,\sigma^V_\numOsv,\zeta_1,...,\zeta_\numCsv\}$.

An evaluation model of $\phi$ is then the possibly infinite set of evaluations such that for all evaluations, the following holds:
\begin{align*}
\forall 1 \leq t \leq N, 1 \leq i \leq \numIs, 1 \leq\ &o \leq \numOs,\\
1 \leq o^V \leq \numOsv, 1 \leq c^V \leq \numCsv&:\\
    \sigma_o(t)& := val(\expr_o)(t)\\
    \sigma^V_{o^V}(t)& \in [-1, 1]\\
    \zeta_{c^V}(t)& \in [-1, 1]\\
    \zeta_{c^V}(t)& = \zeta_{c^V}(t-1) \text{, if } t > 0
\end{align*}
Where $\textit{val}(\expr_o)(t)$ is defined for
\begin{align*}
    \os_o := \expr_o(\is_1,...,\is_\numIs,\quad \os_1,...,&\os_\numOs,\quad \cs_1, ..., \cs_\numCs,\\
    &\csv_1, ..., \csv_\numCsv, \quad \osv_1, ..., \osv_\numOsv)
\end{align*}
with $\cs_j := C_j$ as follows:
\begin{align*}
    \textit{val}(\is_j)(t) &:= \tau_j(t)\\
    \textit{val}(\os_j)(t) &:= \sigma_j(t)\\
    \textit{val}(\cs_j)(t) &:= C_j\\
    \textit{val}(\csv_j)(t) &:= \zeta_j(t)\\
    \textit{val}(\osv_j)(t) &:= \sigma^V_j(t)\\
    \textit{val}(f(\expr_1,...,\expr_k))(t) &:=\\
    f(\textit{val}(\expr_1)(t) &,...,\textit{val}(\expr_k)(t))\\
    \textit{val}(\expr.\textit{offset}(\textit{by: }i,\textit{ or: } l))(t) &:=\\
      \mathclap{\qquad\begin{cases}
            \textit{val}(\expr)(t+i) &\textit{for } 1 \leq t+i \leq N\\
            l &\text{otherwise}
        \end{cases}}
\end{align*}
An \rlola monitor for a specification $\phi$ with the evaluation model $\varphi$ and input streams $\is_1,...,\is_\numIs$ given the uncertain series of measurements $m_1, ..., m_\numIs$ computes a (symbolic) representation of a set of evaluations 
\[\varphi' := \{m_1, ..., m_\numIs,\sigma_1,...,\sigma_\numOs, \sigma^V_1,...,\sigma^V_\numOsv,\zeta_1,...,\zeta_\numCsv\}\]
for any set of output streams $\sigma_1,...,\sigma_\numOs$, streams of slack values $\sigma^V_1,...,\sigma^V_\numOsv$, and constant slack values $\zeta_1,...,\zeta_\numCsv$ such that $\varphi' \subseteq \varphi$.
Triggers are then evaluated existentially based on the set $\psi'$.

\subsection{Boolean Conditions.}
\label{sec:boolean_cond}
With slack variables, stream equations may resolve to symbolically represented intervals instead of scalar values.
We use the ternary predicates $>_p$ and $<_p$ to compare intervals to scalar thresholds in relation to an overlap percentage $p$.
The overlap percentage sets a bound on the overlap of the interval with the threshold such that the predicate still evaluates to true:

The predicate $\expr_i >_{p} c$ is satisfied for $p \in [0,1]$ if the stream expression $\expr_i$ resolves to range $[l, u]$ and $(u-c)/(u-l) > p$ holds.
The definition for $<_p$ is analogous.

The explicit definition of an overlap percentage concretizes the \emph{inconclusive} verdict found in other logics with robust semantics (cf.~\cite{RobustSTL,DBLP:conf/cav/DonzeFM13}). There, comparing an interval of values to a scalar threshold produces an inconclusive verdict if the interval overlaps the threshold.
The $>_p$ and $<_p$ predicates allow for a more precise assessment of the overlap. 

\section{Online Monitoring of \rlola Specifications}\label{sec:online}
We first present an online monitoring algorithm for \rlola that over-approximates the semantics presented above based on interval arithmetic.
We then show a translation for \rlola specifications for which this approximation is indeed precise.

\subsection{Approximate Online Monitoring}
\label{sec:approx_alg}
We first present an online monitoring algorithm for \rlola that over-approximates the semantics presented above.
An evaluation algorithm for \rlola has to manage two potentially unbounded quantities: the number of equations the monitor has to keep in memory and the length of these equations.

The number of equations in memory can be unbounded, as, in general, slack variables can temporally relate measurement errors.
The offline algorithm presented in \Cref{sec:offline} relies on this to refine previous verdicts of the monitor.
However, this requires storing all past monitoring equations for later reevaluation, which results in a memory bound growing with the input trace length.
For online monitoring, we argue that the monitor's constant memory footprint is more important than refining previous verdicts, allowing the monitor to evict old stream equations.

Still, the length of the equations can grow beyond any bound if more and more slack variables accumulate.
Consider the following example:
\begin{lstlisting}
input a_raw: Float
output e: Variable
constant d: Variable

output a := a_raw + 2 * e + 0.5 * d
output sum := sum.last(or: 0) + a
\end{lstlisting}
By definition, the equation for \lstinline{sum} at time 3 includes the slack variables $e_1, e_2$, and $e_3$.
As time progresses, the equation for \lstinline{sum} grows, accumulating more and more slack variables.
One approach to evaluate \rlola specifications is to immediately interpret slack variables as the interval~$[-1,1]$.

\paragraph{Interval Arithmetic}
Interval arithmetic~\cite{IntervalAnalysis} lifts arithmetic operations such as addition and subtraction to intervals.
An uncertain scalar value $x$ can be represented as an interval $[a, b]$ of all possible values that $x$ might have.
An \emph{interval} is defined as a set of real values $[a, b] = \{x \mid a \leq x \leq b\}$.
Arithmetic operations and functions are then defined as follows: For two intervals $a = [a_l, a_u]$ and $b = [b_l, b_u]$ it holds that:
\begin{align*}
    a + b &= [a_l + b_l, a_u + b_u]\\
    a - b &= [a_l - b_u, a_u - b_l]\\
    a * b &= [\text{min}(a_lb_l, a_lb_u, a_ub_l, a_ub_u),\\
          &\qquad\text{max}(a_lb_l, a_lb_u, a_ub_l, a_ub_u)]\\
    a / b &= a * \frac{1}{b} \text{ with } \frac{1}{b} = [{b_u}^{-1}, {b_l}^{-1}], \quad\text{if } 0 \not\in b
\end{align*}
In general, for any monotonic operation $\cdot$ it holds that:
\begin{align*}
    a \cdot b = [&\text{min}(a_l \cdot b_l, a_l \cdot b_u, a_u \cdot b_l, a_u \cdot b_u),\\
        &\text{max}(a_l \cdot b_l, a_l \cdot b_u, a_u \cdot b_l, a_u \cdot b_u)]
\end{align*}

\paragraph{Interval Approximation.}
In the following, we present a monitoring procedure that over-approximates the semantics of the \rlola specification by translating it to a Lola specification defined over intervals using interval arithmetic.
Note that Lola is generic regarding the value domains of streams and their supported operators, which enables this translation.

\begin{definition}[Interval Replacement]
Let $\phi$ be an \rlola specification with input stream variables $\is_1,...,\is_\numIs$, output stream variables $\os_1,...,\os_\numOs$, constants $\cs_1, ..., \cs_\numCs$, constant slack variables $\csv_1,...,\csv_\numCsv$, slack variable streams $\osv_1,...,\osv_\numOsv$ and expressions $\expr_1, ..., \expr_\numOs$.
Let $\phi'$ be a Lola specification with expressions $\expr_1',...,\expr_\numOs'$, where $\expr_i'$ is equal to $\expr_i$ with all references to $\csv_1,...,\csv_\numCsv, \osv_1,...,\osv_\numOsv$ replaced with the interval $[-1, 1]$. 
\end{definition}
Boolean conditions, such as trigger conditions, are evaluated using the ternary operators defined in \Cref{sec:boolean_cond}.
\begin{proposition}
    Let $\phi$ be an \rlola specification and $\phi'$ be the Lola specification obtained from $\phi$ using interval replacement.
    Let $\psi$ be the evaluation model of $\phi$ and $\psi'$ be the evaluation model of $\phi'$.
    If it holds for all sub-expressions of $\phi$ of the form \[\text{if } p \text{ then } \expr_1 \text{ else } \expr_2\] that $p$ does not (transitively) reference any slack variable, then it holds for fixed streams of input data $\tau_1,...,\tau_\numIs$ that if $\{\tau_1,...,\tau_\numIs, \sigma_1,...,\sigma_\numOs, \sigma^V_1,...,\sigma^V_\numOsv,\zeta_1,...,\zeta_\numCsv\} \in \psi$ then $\{\tau_1,...,\tau_\numIs,\sigma_1,...,\sigma_\numOs\} \in \psi'$.
\end{proposition}
The proposition states that \emph{interval replacement} indeed produces a Lola specification that over-approximates an \rlola specification.
Intuitively, intervals reintroduce the aliasing problem, which, as described in \Cref{sec:intro}, results in over-approximating measurement noise.
The additional syntactic requirement on \emph{if} conditions stems from conditionals being non-monotonic functions.

\section{Precise Constant Memory Online Monitoring}
\label{sec:bounded_alg}
We now present a syntactic fragment of \rlola for which constant-memory online monitors exist.
% Concretely, we show that in this fragment, a monitor only has to keep track of a finite set of slack variables by pruning linearly dependent expressions of slack variables.
We develop this result in multiple steps.
First, we define two requirements that all specifications in the fragment must fulfill and show how slack variables can be pruned from the monitor if their coefficients are linearly dependent.
Then, we give examples of increasing complexity, highlighting how these requirements ensure the co-linearity of subsets of the slack variables.

\paragraph{Requirement 1.} First, we syntactically limit stream expressions.
They are required to be in one of the following two forms:
\begin{align}
    \expr_i &:= c_s *\os_i[o, d] + c_i^T \begin{pmatrix}\is_1\\\vdots\\\is_\numIs\end{pmatrix} + c_\epsilon^T \begin{pmatrix}\osv_1\\\vdots\\\osv_\numOsv\end{pmatrix} + c_\delta^T \begin{pmatrix}\csv_1\\\vdots\\\csv_\numCsv\end{pmatrix}\\
    \expr_i &:= \text{if } p \text{ then } \expr_i^c \text{ else } \expr_i^a
\end{align}
Where $c_s \in \{0, 1\}$, $p$ is a boolean stream expression and $c_i \in \Real^\numIs, c_\epsilon \in \Real^\numOsv, c_\delta \in \Real^\numCsv$.
% $c^i_1, ..., c^i_\numIs, c^\epsilon_1, ..., c^\epsilon_\numOsv, c^\delta_1, ..., c^\delta_\numCsv \in \Real$.
Intuitively, this requirement ensures that at every point in time, each equation entailed by a stream expression is an affine form \cite{AffineArithmetic}.

\paragraph{Requirement 2.} Second, we require that every output stream only occurs in dependency loops of equal weight.    
Concretely, let $\phi$ be a specification with output streams $\os_1, ..., \os_\numOs$ and let $G = (V, E)$ be the dependency graph of $\phi$, then:
\begin{align*}
    \forall 1 \leq i \leq \numOs. &\exists c_o \in \mathbb{Z}.\\
    &\forall \langle\os_i \xrightarrow{o_1} ... \xrightarrow{o_n} \os_i\rangle \in E. \sum_{1 \leq r \leq n}o_r = c_o
\end{align*}
Intuitively, this requirement ensures that equations in the equation store are affine.
For example, it prohibits calculating the Fibonacci sequence.

One exception from these requirements are trigger streams.
As their value, by definition, cannot be used by other streams, they can express arbitrary boolean assessments over output streams.

Consider \Cref{ex:motivating}; If the \lstinline{distance} output stream is inlined, all stream expressions in the specification satisfy Requirement 1.
Requirement 2 is also satisfied, as \lstinline{position_x} and \lstinline{position_y} are both only part of self-loops with weight -1.

We now develop the construction of constant-memory Lola monitors that monitor specifications of the fragment without loss of precision.
The key idea is that we do not track \emph{when} uncertainty occurs, but only how much uncertainty accumulates in each equivalence class of behavior.
Instead of introducing a fresh slack variable for each time step, we represent all occurrences of a slack source by a single constant slack variable.
The temporal accumulation of uncertainty is then captured by counting how often this variable contributes to a stream and scaling it accordingly.

For this, we first define a method to reduce the number of slack variables in equations.
\begin{definition}[Slack Variable Pruning]
\label{def:pruning}
    Let $\vec{\epsilon} \in [-1,1]^j$ be a vector of slack variables and let $C \in \Real^{k \times j}$ be a matrix of coefficients, then
    \[
    y = C (\epsilon_1, ...,\epsilon_j)^T
    \] 
    defines a zonotope over slack variables $\epsilon_1,...,\epsilon_j$.
    To prune slack variables, we reduce the dimension of $\Vec{\epsilon}$ by finding collinear column vectors of the matrix $C$ to obtain an equivalent (pruned) representation of the zonotope:
    \[
    y = C'(\epsilon_1, ...,\epsilon_l)^T
    \]
    with $C' \in \Real^{k \times l}$ for $l \leq j$.
    If two or more column vectors of $C$ are collinear, it holds that $l < j$.
\end{definition}
Intuitively, collinear columns in $C$ correspond to slack variables that always influence the monitored values in the same direction and with a fixed ratio. Hence, their joint effect can be represented by a single slack variable whose coefficient captures the combined magnitude.

We use \Cref{def:pruning} to prune variables from the equations the monitor maintains at runtime.
For that, we define the state of a monitor as follows:

\paragraph{Monitoring State.}
Let $\phi$ be an \rlola specification with output streams:
\[
    \os_1 := \expr_1 \quad ... \quad \os_\numOs := \expr_\numOs
\]
A monitor manages two sets of equations called equation stores: $R$ for resolved equations of the form: $\sigma_i(t) = c' + c_1 \epsilon_1 + \dots + c_s \epsilon_s$ and $U$ for unresolved equations of the form $\sigma_i(t) = \expr_i$. At time $t$ for measurements $m_1,...,m_\numIs$ the following equations are added to $R$: $\tau_1(t) = m_1, ..., \tau_I(t) = m_I$ and $\sigma_1(t) = \expr_1, ..., \sigma_\numOs(t) = \expr_\numOs$ to $U$.
If, through simplifications, equations from $U$ become resolved, they are moved to $R$.

At any time point $t$, the sets $U$ and $R$ are called the monitoring state. We call the equations in $R$ monitoring equations.
\subsection{Translating Different Coefficients}
As a first example, consider the specification in \Cref{fig:different_coeff} and its partial monitoring state depicted below:
\begin{example}[An \rlola specification where a slack variable occurs with different coefficients.]\label{fig:different_coeff}
\begin{lstlisting}
        input a_raw: Float
        
        output e: Variable
        constant d: Variable
        
        output a := a_raw + e + d
        output sum2 := sum2.last(or: 0) + 2a
        output sum3 :=  sum3.last(or: 0) + 3a
\end{lstlisting}
\begin{center}
\bgroup
\footnotesize
\def\arraystretch{1.5}%
\setlength\tabcolsep{0.4em}
\begin{tabular}{l|rrrr}
       & 1 & 2 & 3\\
     \hline
     sum2 & $2e_1 + 2d$ & $2e_1 + 2e_2 + 4d$ & $2e_1 + 2e_2 + 2e_3 + 6d$\\
     sum3 & $3e_1 + 3d$ & $3e_1 + 3e_2 + 6d$ & $3e_1 + 3e_2 + 3e_3 + 9d$\\
\end{tabular}
\egroup
\end{center}
\end{example}
The table depicts partial monitoring equations truncated to their slack variable part at time points one to three.
As discussed in \Cref{sec:approx_alg}, one can see that the slack variables produced by \lstinline{e} accumulate in the monitoring equations of \lstinline{sum2} and \lstinline{sum3}.
Consider the equations at time three in their vector representation, omitting the constant slack variable $d$:
\[
\begin{pmatrix}
    2, 2, 2\\
    3, 3, 3
\end{pmatrix}
(e_1,e_2,e_3)^T
\]
All column vectors of the matrix are identical, hence collinear, and can be pruned as defined in \Cref{def:pruning}.
In fact, this holds for every time step due to Requirement 1 which ensures that slack variables only occur with a constant coefficient in stream equations.
Therefore, the contribution of $(e_1,\dots,e_3)$ can be rewritten as
\[
\begin{pmatrix}
    2\\
    3
\end{pmatrix} \cdot (e_1 + e_2 + e_3)
\]
Since each $e_i \in [-1,1]$, their sum ranges over an interval that can be represented by a single slack variable scaled by the number of occurrences. Hence, instead of tracking each $e_i$ separately, it suffices to track how often the slack variable occurs and multiply a single representative variable accordingly.

Based on this, \Cref{fig:different_coeff:translated} depicts an equivalent \rlola specification using only constant slack variables.
\begin{lstlisting}[caption={The translated \rlola specification from \Cref{fig:different_coeff}}, label=fig:different_coeff:translated]
    input a_raw: Float
    constant e: Variable
    constant d: Variable
    
    output e_coeff :=
      e_coeff.last(or: 0) + 1
    output sum2_raw := 
      sum2_raw.last(or: 0) + 2 * a_raw + 2d
    output sum3_raw := 
      sum3_raw.last(or: 0) + 3 * a_raw + 3d
    output sum_2 := sum2_raw + 2 * e_coeff * e
    output sum_3 := sum3_raw + 3 * e_coeff * e
\end{lstlisting}
The stream \lstinline{e_coeff} counts the number of occurrences, while the constant slack variable \lstinline{e} represents the aggregated uncertainty.

\subsection{Translating Different Offsets}
Next, consider the example where slack variables are used in streams that reference themselves with different offsets in \Cref{fig:different_offsets}.
Note that the constant slack variable $d$ is omitted for simplicity.
\begin{example}[A specification where slack variables accumulate under different offsets.]\label{fig:different_offsets}
    \begin{lstlisting}
        input a_raw: Float
        output e: Variable
        
        output a := a_raw + e
        output sum := sum.offset(by: -1, or: 0) + 2a
        output eo_sum := eo_summ.offset(by: -2, or: 0) + 3a
    \end{lstlisting}
\end{example}
Again, consider the partial monitoring equations for this specification in the table below:
\begin{center}
\bgroup
\footnotesize
\def\arraystretch{1.5}%
\setlength\tabcolsep{0.4em}
\begin{tabular}{l|rrrr}
       & 1 & 2 & 3 & 4\\
     \hline
     sum & $2e_1$ & $2e_1 + 2e_2$ & $2e_1 + 2e_2 + 2e_3$ & $2e_1 + 2e_2 + 2e_3 + 2e_4$\\
     eo\_sum & $3e_1$ & $3e_2$ & $3e_1 + 3e_3$ & $3e_2 + 3e_4$\\
\end{tabular}
\egroup
\end{center}
Because of the offset of $-2$, a slack variable $e_i$ is added at either an even position \emph{or} an odd position of \lstinline{eo_sum}, never at both.
This stems from the stream-based semantics of \rlola.
Because of this, we analyze these two cases separately.
Consider the equations at time three and four in their vector representation:\\
\begin{minipage}{0.45\linewidth}
\[
\begin{pmatrix}
    2, 2, 2\\
    3, 0, 3
\end{pmatrix}
(e_1,e_2,e_3)^T
\]
\end{minipage}
\hfill
\begin{minipage}{0.45\linewidth}
\[
\begin{pmatrix}
    2, 2, 2, 2\\
    0, 3, 0, 3
\end{pmatrix}
(e_1,e_2,e_3, e_4)^T
\]
\end{minipage}
\\

\noindent Both matrices can be pruned to reduce the number of slack variables to two.
With the same argument as for different coefficients, this holds for all even and odd positions, respectively.
The separation into even and odd positions ensures that, within each case, slack variables again appear with identical coefficient vectors. Consequently, pruning can be applied independently in each case, yielding one slack variable per case without loss of precision.

We give an equivalent \rlola specification that only uses constant slack variables in \Cref{fig:different_offsets:translated}.
\begin{lstlisting}[caption={The translated \rlola specification from \Cref{fig:different_offsets}}, label=fig:different_offsets:translated]
        input a_raw: Float
        constant e_even: Variable
        constant e_odd: Variable
        
        output step := step.last(or: 0) + 1
        output e_even_coeff := if step % 2 = 0 $\land$ step % 1 = 0
                               then e_even_coeff.last(or: 0) + 1 else e_even_coeff.last(or: 0)
        output e_odd_coeff := if step % 2 = 1 $\land$ step % 1 = 0
                              then e_odd_coeff.last(or: 0) + 1 else e_odd_coeff.last(or: 0)
        
        output sum_raw := sum_raw.last(or: 0) + a_raw
        output eo_sum_raw := eo_sum_raw.offset(by: -2, or: 0) + a_raw
        output sum := sum_raw + 2 * e_even_coeff * e_even + 2 * e_odd_coeff * e_odd
        output eo_sum := if step % 2 = 0 $\land$ step % 1 = 0
                         then eo_sum_raw + 3 * e_even_coeff * e_even
                         else eo_sum_raw + 3 * e_odd_coeff * e_odd
\end{lstlisting}
Note that the if conditions in the specification can be simplified.
Yet, they are kept as is to demonstrate how the construction scales to arbitrary offsets in multiple streams as long as Requirement 2 is satisfied.
In general, if a slack variable appears in multiple streams with different offsets $o_1, ..., o_k$, the above case distinction has to be extended to $s = o_1 * ... * o_k$ cases of offset combinations resulting in $s$ constant slack variables.

\paragraph{Central Observation.}
When partitioning the monitoring equations of a specification in the fragment by case, the coefficients of the slack variables occurring in each case will always be equal.
This implies that, within each case, all slack variables contribute along the same direction in the value space. By \Cref{def:pruning}, such contributions can be merged into a single slack variable without changing the set of possible valuations. Therefore, one constant slack variable per case suffices to represent all uncertainty precisely.
For example, in the above specification, the coefficients at even and at odd positions will always be equal.
This is explained by the constant coefficients asserted by Requirement 1.
\begin{figure*}[h]
\centering
\begin{tabular}{|c|}\hline \\
\ \ \begin{minipage}{11.6cm}
    Let $C_1,...,C_k$ be the cases in which subsets of slack variables occur with equal coefficients in the monitoring equations.
    \begin{enumerate}
        \item For each case $C_i$ introduce a constant slack variable $d_i$ and construct a stream that counts how often this case occurs as follows:
        \begin{lstlisting}
            constant $d_i$: Variable
            output $c_i$ := if $C_i$ then $c_i$.offset(by:-1, or:0) + 1 else $c_i$.offset(by:-1, or:0)
        \end{lstlisting}
        \item For each output stream $s$ and slack variable stream $\epsilon$, where $s$ references $\epsilon$ in cases $C_i,...,C_j$ with coefficients $x_i, ..., x_j$ add a stream that reconstructs $\epsilon$ for $s$:
        \begin{lstlisting}
            output s_$\epsilon$ := $c_i$ * $x_i$ * $d_i$ + ... +  $c_j$ * $x_j$ * $d_j$
        \end{lstlisting}
        \item For each output stream $s$ and constant slack variable $\delta$, where $s$ references $\delta$ in cases $C_i,...,C_j$ with coefficients $x_i, ..., x_j$ add a stream that reconstructs $\delta$ for $s$:
        \begin{lstlisting}
            output s_$\delta$:= ($c_i$ * $x_i$ + ... +  $c_j$ * $x_j$) * $\delta$
        \end{lstlisting}
        \item For each output stream $s$ that references slack variable streams $\epsilon_1, ..., \epsilon_n$ and constant slack variables $\delta_1,...,\delta_m$ construct a stream $s\_raw$ that is equal to $s$, apart that all references to slack-variables are removed.
        Construct a stream $s'$ that reconstructs $s$ from its partial sums.
        \begin{lstlisting}
            output $s'$ := s_raw + s_$\epsilon_1$ + ... + s_$\epsilon_n$ + s_$\delta_1$ + ... + s_$\delta_m$
        \end{lstlisting}
    \end{enumerate}
    \end{minipage}\ \ \  \\ \hline
    \end{tabular}
    \caption{Construction of constant-memory monitors.}
    \label{fig:bounded_construction}
\end{figure*}
\subsection{Translating \emph{If} Clauses}
Lastly, we extend this case distinction to \emph{if} clauses. 
Consider the following example:
\begin{example}[A specification where slack variables accumulate under an if clause.]
\begin{lstlisting}
input a_raw: Float

output e: Variable
output a := a_raw + e
output sum := if a_raw > 10 
              then sum.offset(by: -1, or: 0) + 2a
              else sum.offset(by: -1, or: 0) + 5a
output eo_sum :=  eo_sum.offset(by: -2, or: 0) + 3a
\end{lstlisting} 
\end{example}
To handle \emph{if} clauses, we distinguish one case per if condition.
Let a slack variable output stream be referenced in $n$ output streams that contain \emph{if} conditions.
Let there be a total of $k$ if conditions in their stream expressions. 
Then, a monitor has to handle $s * (n+k)$ slack variables where $s$ is the previous bound on the number of slack variables.

In the above example, there are two different offsets (-1 and -2), and one stream contains a total of one if condition.
Hence, to precisely monitor the above specification, the monitor has to distinguish $2 * (1+1) = 4$ cases.
In the following, we group the coefficients by their case:
\begin{center}
\bgroup
\footnotesize
\def\arraystretch{1}%
\setlength\tabcolsep{0.4em}
\begin{tabular}{r|c|c}
& even & odd \\
\hline
$\text{a\_raw} \leq 10$
& $\begin{pmatrix}5\\3\end{pmatrix}e_2$ & $\begin{pmatrix}5\\3\end{pmatrix}e_1 + \begin{pmatrix}5\\3\end{pmatrix}e_5$\\
$\text{a\_raw} > 10$ & $\begin{pmatrix}2\\3\end{pmatrix}e_4$ & $\begin{pmatrix}2\\3\end{pmatrix}e_3$\\
\end{tabular}
\egroup
\end{center}
By \Cref{def:pruning}, the variables $e_1$ and $e_5$ of the above example can be pruned.

Following the central observation, it is easy to see that a monitor for the fragment only needs to keep a \emph{single} constant slack variable for each case for each stream of slack variables. \Cref{fig:bounded_construction} summarizes the construction.

The specifications generated by this construction can be evaluated, without any loss of precision, using the algorithm from \Cref{sec:approx_alg}.
This is because all constant slack variables generated by the construction are only referenced once in streams that are \emph{state-less}, meaning that they do not propagate through computations, preventing the aliasing problem.
\begin{proposition}
    The construction in \Cref{fig:bounded_construction} is correct and only requires a bounded number of slack variables.
\end{proposition}
As the specification is finite, there can only be a bounded number of cases; hence, the number of slack variables is also bounded.
Correctness follows from two observations:
\begin{enumerate}
    \item By \Cref{def:pruning}, replacing collinear slack variables with a single variable preserves the set of possible valuations of each monitoring equation.
    \item By the requirements of the fragment, at each time step a slack variable contributes to exactly one case per stream, and within each case all coefficients are constant.
\end{enumerate}
Therefore, the construction preserves the semantics of the original specification while reducing the number of slack variables to a bounded set.

The online construction shows that uncertainty can be represented compactly by aggregating slack variables into a bounded set of constant slack variables, enabling precise monitoring with fixed memory.
We now turn to an alternative setting that removes these constraints and allows for more global reasoning.
\section{Offline Monitoring of \rlola Specifications}\label{sec:offline}
In contrast to online monitoring, which must produce immediate verdicts based on local information at each time step, offline monitoring operates on a finite trace and enables reasoning over the entire execution.

Instead of relying on overlap percentages, offline monitoring evaluates trigger conditions over the set of ground truth traces that are compatible with the observed measurements. Compatibility, as defined in \Cref{sec:error-model}, requires the existence of an assignment to slack variables that explains the measurements. As a result, constraints that appear locally consistent may become contradictory when considered across multiple time steps, as discussed in~\Cref{sec:motivating}.

This section presents the SMT-based offline analysis algorithm for \rlola.
We begin by relating the interpretation of trigger conditions in offline monitoring to the interpretation in online monitoring.

\subsection{Boolean Trigger Conditions in Online and Offline  Monitoring}
Besides the constant memory requirement, online monitoring algorithms face another key requirement:
It must provide an immediate verdict, as the system may enter a dangerous or unsafe state precisely at the moment of evaluation.

To meet this requirement, we propose the ternary predicates $>_p$ and $<_p$ in \Cref{sec:boolean_cond}.
These predicates compare symbolically represented intervals to scalar thresholds based on a specified overlap percentage $p$, avoiding inconclusive verdicts during runtime.

In offline monitoring, the monitor does not operate in real time alongside the system.
Consequently, the constraints associated with real-time operation, such as constant memory usage and immediate verdicts, no longer apply, as the input data is pre-recorded and has a finite length.
This relaxation of requirements allows offline monitoring to provide valuable insights through post-execution analysis, enabling a more precise evaluation of the system's execution than is possible during runtime.

Since the above real-time requirements do not apply to offline monitoring, there is no need for ternary predicates in trigger conditions. Instead, triggers can be evaluated either existentially or universally over the entire execution trace.
Specifically, a specification is considered satisfied if there exists a ground truth trace consistent with the measurements such that no trigger evaluates to true.

Formally, a specification $\phi$ with the evaluation model $\varphi$, input streams $\is_1,...,\is_\numIs$ and trigger streams $\os_i, \dots, \os_j$  is satisfied by a series of measurements $m_1,...,m_\numIs$ if:

\begin{align*}
    \exists \varphi' &\in \varphi.\\
    &\varphi' = \{m_1, ..., m_\numIs,\sigma_1,...,\sigma_\numOs, \sigma^V_1,...,\sigma^V_\numOsv,\zeta_1,...,\zeta_\numCsv\}\\
    &\land \sigma_i = \overrightarrow{\bot} \land \dots \land \os_j = \overrightarrow{\bot}
\end{align*}
Where $\overrightarrow{\bot}$ is the constant false stream.
Using this definition of satisfiability, we present an SMT encoding of specifications for offline analysis.

\subsection{An SMT Encoding for \rlola Specifications}\label{sec:encoding}
We now present an SMT encoding of \rlola specifications that evaluates trigger conditions existentially over a trace, catching violations that an online analysis cannot detect.
For this, we first fix a specification $\phi$ and a system execution trace $\Theta$.
Intuitively, the encoding then represents an unrolling of the semantics presented in \Cref{sec:semantics} over the trace $\tau$.

Let the specification $\phi$ contain the constants $\cs_1, ..., \cs_\numCs$, input streams $\is_1,...,\is_\numIs$, output streams $\os_1,...,\os_\numOs$, of which $\trs_1,...,\trs_\numTs$ are trigger streams, constant slack variables $\csv_1,...,\csv_\numCsv$, and slack variable streams $\osv_1,...,\osv_\numOsv$.
Let $\Theta$ contain the series of measurements $\tau_1,...,\tau_\numIs$ of length $N$.
We use $\tau_i[t]$ to reference a single measurement of the series at time $1 \leq t \leq N$.
We encode the positions of streams as separate variables. 
Consequently, input, output, and slack variable streams are encoded using $N$ variables, while constants and constant slack variables require only one variable.
Similarly, we refer to the SMT variable corresponding to the value of stream $s$ at time $1 \leq t \leq N$ as $s[t]$.
In the following, we construct the conjunctive constraint system for the specification:

\newcommand{\constr}{\textit{constr}}
\begin{align*}
    \mathrlap{\hspace{-6em}\text{A \emph{constant} $\cs := C$ in the specification is translated to the}}\\
    \mathrlap{\hspace{-6em}\text{constraint:\hspace{10em}}}\\
    &\cs = C\\
    \mathrlap{\hspace{-6em}\text{A \emph{constant slack variable} $\csv$ is encoded as:}}\\
    &\csv \leq 1 \land \csv \geq -1\\
    \mathrlap{\hspace{-6em}\text{A \emph{slack variable stream} $\osv$ is translated to:}}\\
    &\bigwedge^N_{t = 1} \osv[t] \leq 1 \land \osv[t] \geq -1\\
    \mathrlap{\hspace{-6em}\text{An \emph{input stream} $\is_i$ is assigned its measurement as:}}\\
    &\bigwedge^N_{t = 1} \is_i[t] = \tau_i[t]
\end{align*}
\begin{align*}
    \mathrlap{\hspace{-6em}\text{An \emph{output stream} $\os := \expr$ generates the constraint:}}\\
    &\bigwedge^N_{t = 1} \os[t] = \constr(t, expr)\\
    \mathrlap{\hspace{-6em}\text{A \emph{trigger} condition $\expr$ translates to:}}\\
    &\bigwedge^N_{t = 1} \neg\constr(t, expr)
\end{align*}
Where $\constr(t, \expr)$ is defined as:
\begin{align*}
    \constr(t, \cs) &:= \cs\\
    \constr(t, \csv) &:= \csv\\
    \constr(t, \is) &:= \is[t]\\
    \constr(t, \os) &:= \os[t]\\
    \constr(t, \osv) &:= \osv[t]\\
    \constr(t, f(\expr_1,...,\expr_k)) &:=\\
    f(\constr(t, \expr_1) &,...,\constr(t, \expr_k))\\
    \constr(t, \expr.\textit{offset}(\textit{by: }i,\textit{ or: } \expr_d)) &:=\\
        \mathclap{\qquad\begin{cases}
            \constr(t+i, \expr) &\textit{for } 1 \leq t+i \leq N\\
            \constr(t, \expr_d) &\text{otherwise}
        \end{cases}}
\end{align*}
It follows that the trace $\theta$ satisfies the specification $\phi$, iff the constraint system is satisfiable.

\subsection{An Example Constraint System}
In the following, we present the SMT encoding of the specification shown in \Cref{ex:motivating} in the SMT-Lib format\cite{barrett2010smt}.
For simplicity, we only included the first two points of the trace depicted in \Cref{fig:running_example}.
\vspace{1em}
\begin{lstlisting}[language=SMT]
; Direction Input Stream Variables
(declare-fun direction0 () Int)
(declare-fun direction1 () Int)

; Raw Distance Input Stream Variables
(declare-fun raw_distance0 () Real)
(declare-fun raw_distance1 () Real)

; Delta Slack Variable
(declare-fun delta () Real)

; Epsilon Slack Variables
(declare-fun epsilon0 () Real)
(declare-fun epsilon1 () Real)

; Distance Output Stream Variables
(declare-fun distance0 () Real)
(declare-fun distance1 () Real)

; Position X Output Stream Variables
(declare-fun position_x0 () Real)
(declare-fun position_x1 () Real)

; Position Y Output Stream Variables
(declare-fun position_y0 () Real)
(declare-fun position_y1 () Real)

; Direction Measurements
(assert (= direction0 3))
(assert (= direction1 3))

; Raw Distance Measurements
(assert (= raw_distance0 (/ 1131.0 1000.0)))
(assert (= raw_distance1 (/ 879.0 500.0)))

; Constrain Slack Variables to [-1, 1]
(assert (>= delta (- 1.0)))
(assert (<= delta 1.0))

(assert (>= epsilon0 (- 1.0)))
(assert (<= epsilon0 1.0))

(assert (>= epsilon1 (- 1.0)))
(assert (<= epsilon1 1.0))

; Distance Output Stream Equations
(assert (= distance0 (+ raw_distance0 (* (/ 1.0 10.0) epsilon0) (* (/ 9.0 10.0) delta))))
(assert (= distance1 (+ raw_distance1 (* (/ 1.0 10.0) epsilon1) (* (/ 9.0 10.0) delta))))

; Postion X Stream Equation Step 0
(assert (=> (= direction0 1) 
   (= position_x0 distance0)))
(assert (=> (= direction0 (- 1)) 
   (= position_x0 (- distance0))))
(assert (=> (= direction0 3)
    (= position_x0 (* (/ 1767766952966369.0 2500000000000000.0) distance0))))
(assert (=> (= direction0 (- 3))
    (= position_x0
       (* (- (/ 1767766952966369.0 2500000000000000.0)) distance0))))
(assert (=> (= direction0 4)
    (= position_x0
       (* (- (/ 1767766952966369.0 2500000000000000.0)) distance0))))
(assert (=> (= direction0 (- 4))
    (= position_x0 (* (/ 1767766952966369.0 2500000000000000.0) distance0))))
(assert (=> (and (distinct direction0 1)
                 (distinct direction0 (- 1))
                 (distinct direction0 3)
                 (distinct direction0 (- 3))
                 (distinct direction0 4)
                 (distinct direction0 (- 4)))
    (= position_x0 0.0)))

...

; Postion Y Stream Equation Step 1
(assert (=> (= direction1 2) 
    (= position_y1 (+ position_y0 distance1))))
(assert (=> (= direction1 (- 2)) 
    (= position_y1 (- position_y0 distance1))))
(assert (=> (= direction1 3)
    (= position_y1
       (+ position_y0
          (* (/ 1767766952966369.0 2500000000000000.0) distance1)))))
(assert (=> (= direction1 (- 3))
    (= position_y1
       (- position_y0
          (* (/ 1767766952966369.0 2500000000000000.0) distance1)))))
(assert (=> (= direction1 4)
    (= position_y1
       (+ position_y0
          (* (/ 1767766952966369.0 2500000000000000.0) distance1)))))
(assert (=> (= direction1 (- 4))
    (= position_y1
       (- position_y0
          (* (/ 1767766952966369.0 2500000000000000.0) distance1)))))
(assert (=> (and (distinct direction1 2)
                 (distinct direction1 (- 2))
                 (distinct direction1 3)
                 (distinct direction1 (- 3))
                 (distinct direction1 4)
                 (distinct direction1 (- 4)))
    (= position_y1 position_y0)))

; Negated Trigger Conditions
(assert (>= position_x0 1.0))
(assert (<= position_x0 5.0))

(assert (>= position_y0 1.0))
(assert (<= position_y0 5.0))

(assert (>= position_x1 1.0))
(assert (<= position_x1 5.0))

(assert (>= position_y1 1.0))
(assert (<= position_y1 5.0))

(check-sat)
\end{lstlisting}
The necessary variable definitions at the top of the encoding are followed by assigning the measurements to the input stream variables.
Afterward, slack variables are constrained to their range from $[-1, 1]$.
The arithmetic operations in the output streams are encoded as expected in the SMT formula, while the conditional clauses $\texttt{if } b \texttt{ then } \expr_1 \texttt{ else } \expr_2$ translates to (nested) implications of the form $b \implies \expr_1 \land  \neg b \implies \expr_2$.
Lastly, the negated trigger conditions are asserted, and the solver is instructed to check for the constraint system's satisfiability.

\section{Tool Overview}\label{sec:tool}
We implemented the presented algorithms in the RTLola framework~\cite{frontend,interpreter}, written in Rust.
The framework consists of a frontend that parses and analyzes specifications, and an interpreter backend that evaluates them.
The frontend performs basic analysis steps, such as type and naming checks, and constructs the dependency graph of the specification.
For this work, we extended the analyses to support slack variables and affine forms and added automatic checks for Requirements one and two from \Cref{sec:bounded_alg}.
If all analyses succeed, the frontend exports the specification to a condensed, machine-readable intermediate representation for the backend.

The interpreter consumes this intermediate representation and evaluates the specification on a given trace.
We extended it to support intervals and affine forms as stream values and implemented the corresponding arithmetic primitives.
For offline analysis, the interpreter can either export the constraint system to a file or invoke the SMT solver Z3~\cite{DBLP:conf/tacas/MouraB08} directly to check satisfiability, using the Rust bindings for Z3 to build the constraint system.

\paragraph{Specification Analysis.}
The frontend parses a textual specification into an abstract syntax tree (AST).
On the AST, it performs several analyses, including naming checks to ensure that all referenced streams are declared and type checks to validate that all operations use consistent data types.
It constructs the dependency graph defined in \Cref{sec:Lola}, checks well-formedness, and verifies requirement two.
Requirement one is verified by an additional AST traversal that checks the corresponding syntactic condition.
To improve accessibility, we implemented informative error messages that guide users in designing specifications that satisfy the requirements.
The following example illustrates a violation of requirement two:
\begin{Verbatim}[commandchars=\\\{\},fontsize=\small]
\textcolor{red}{error}: \textbf{Only equal weight cycles are allowed in}
\textcolor{teal}{  ¦}    \textbf{affine specifications.}
\textcolor{teal}{  ¦}    
\textcolor{teal}{6 ¦}\textcolor{errorred}{ + + output y := if x > 5.0 then}
\textcolor{teal}{7 ¦}\textcolor{errorred}{ ¦ ¦     y.offset(by: -1).defaults(to: 0.0)}
\textcolor{teal}{8 ¦}\textcolor{errorred}{ ¦ ¦ else}
\textcolor{teal}{9 ¦}\textcolor{errorred}{ ¦ ¦     x.offset(by: -1).defaults(to: 0.0)}
\textcolor{teal}{  ¦}\textcolor{errorred}{ ¦ ¦        + a + 0.5 * d + 3.0 * e}
\textcolor{teal}{  ¦}\textcolor{errorred}{ ¦ ¦}
\textcolor{teal}{  ¦}\textcolor{errorred}{ +-¦ Found cycle y -> y with weight -1.}
\textcolor{teal}{  ¦}\textcolor{errorred}{   + Found other cycle y -> x -> y with weight -2.}
\end{Verbatim}

\paragraph{Specification Interpretation.}
To evaluate specifications, we implemented affine forms and their pruning directly in the interpreter rather than relying on specification-to-specification translation, in order to improve runtime performance.
The interpreter represents affine forms as row vectors containing the coefficients of the contributing slack variables.
For pruning, it aggregates all affine forms into a matrix $A$ representing the affine state of the monitor.
It then searches for collinear column vectors by computing the Euclidean norm of $A$ as $|A|$ and the pairwise dot products of column vectors as $R = |A|^T \cdot |A|$.
By definition of the dot product, if the entry in row $i$ and column $j$ of $R$ is $1$, then the $i_\textit{th}$ and $j_\textit{th}$ slack variables can be pruned.
For specifications in the precise fragment defined in this paper, this method eventually identifies two collinear column vectors.

\paragraph{Input and Output Format.}
The frontend parses textual specifications.
The interpreter provides both an API and a CLI interface~\cite{cav4,DBLP:conf/fm/BaumeisterFKS24}.
For offline analysis, traces can be imported from CSV files with one column per input stream.
For online analysis, inputs can be read from the application’s stdin in CSV format.
By default, the CLI outputs verdicts in a human-readable log format.
Output stream values can also be exported to a CSV file or accessed through the API to construct user-defined data structures.

\section{Evaluation}\label{sec:eval}
\begin{figure*}[ht]
\centering  
\begin{subfigure}{0.49\linewidth}
    \includegraphics[width=\textwidth]{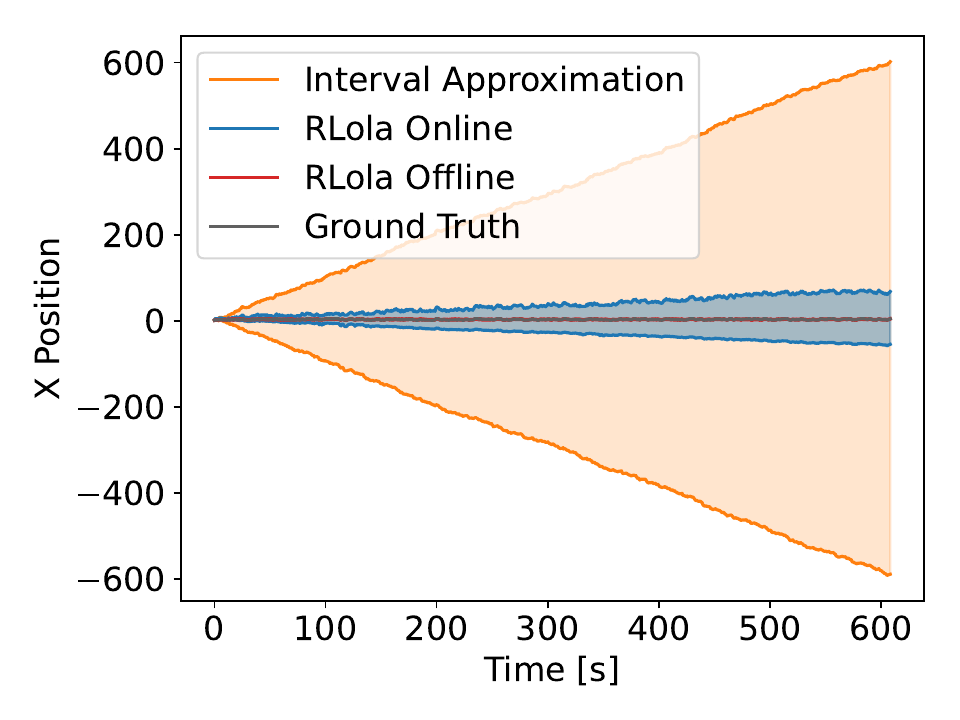}
    \caption{The $x$-position range computed by the interval approximation, the precise online algorithm, and the SMT Encoding.}
    \label{fig:precision_all}
\end{subfigure}
\hfill
\begin{subfigure}{0.49\linewidth}
    \includegraphics[width=\textwidth]{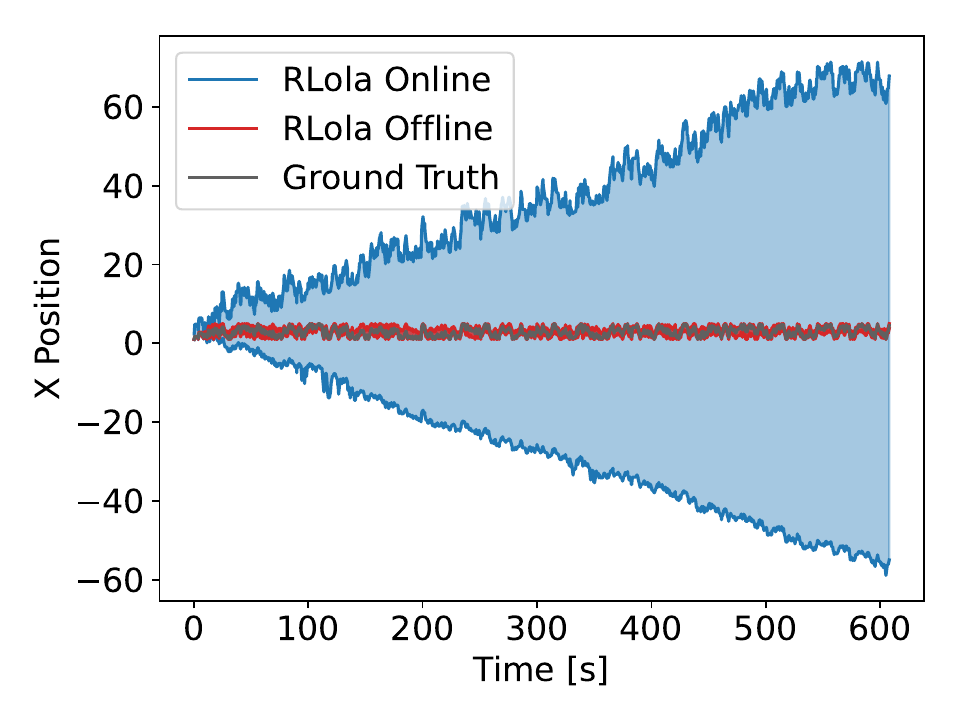}
    \caption{The $x$-position range computed by the precise online algorithm and the SMT Encoding.}
    \label{fig:eval_precision_no_intervals}
\end{subfigure}
\caption{A comparison between $x$-Position ranges used to assess the trigger conditions by the different algorithms.}
\label{fig:eval_precision}
\end{figure*}
In this section, we evaluate the runtime and precision of the presented tool using the running example from \Cref{sec:motivating} and a real-world case study on monitoring vehicle emissions~\cite{DBLP:conf/tacas/BiewerFHKSS21}.
The specification from \Cref{ex:motivating} is analyzed with three algorithms: the interval approximation (\Cref{sec:approx_alg}), the precise online algorithm (\Cref{sec:bounded_alg}), and the offline algorithm (\Cref{sec:encoding}).

For the real-world case study, we use the precise online and offline algorithms. The trace is derived from sensor data provided in~\cite{DBLP:conf/tacas/BiewerFHKSS21}. Following~\cite{DBLP:journals/infsof/HiplerKLMSW26}, we assess the monitor’s precision under varying noise conditions by adjusting both noise amplitude and frequency, and checking whether the expected result is still computed.

All experiments were conducted on a machine with an AMD Ryzen 7 5800X 8-Core processor (4.6GHz) and 32GB of memory. Reported runtimes are averaged over five runs.

\paragraph{Running Example.}
We compare the evaluation algorithms for \rlola using the examples from \Cref{sec:motivating}.
As a baseline, we generate a ground truth trace that satisfies the specification and add noise according to the error model in \Cref{sec:error-model}.

The graphs in \Cref{fig:eval_precision} illustrate the monitor's precision based on the error margin around the robot's x position. \Cref{fig:precision_all} presents the results for all algorithms mentioned above, whereas \Cref{fig:eval_precision_no_intervals} excludes the interval approximation to provide a more focused view of the data.
While the error ranges grow over time, the graph visualizes the pessimistic over-approximation of interval arithmetic caused by the aliasing problem.

This aliasing arises from the constant error \lstinline{delta}.
In \rlola, this error is subtracted from the position when the robot moves in opposite directions, whereas interval analysis adds it in both directions.

\Cref{fig:eval_precision_no_intervals} further highlights the difference between offline and online analyses.
The online algorithm resolves overlaps between error margins and thresholds using percentages and bases its decisions on the blue region.
In contrast, the offline algorithm explicitly constrains positions to satisfy the threshold and tracks only valid positions, shown as the red region.
The specification is considered satisfied as long as this region is non-empty.

\begin{figure*}
\centering
\begin{subfigure}{0.49\linewidth}
    \includegraphics[width=\textwidth]{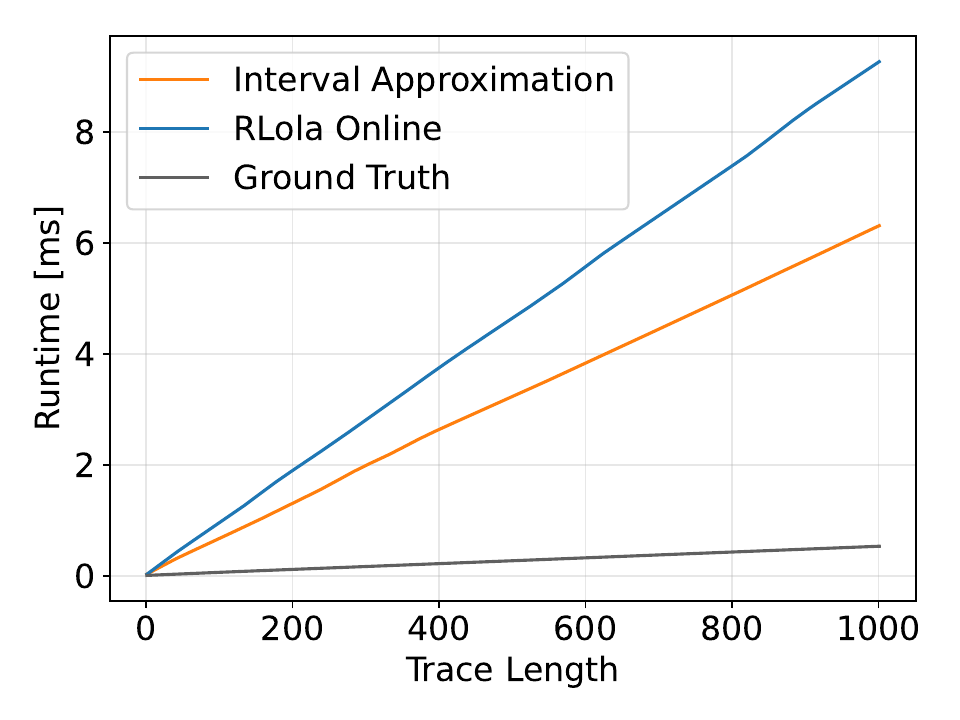}
    \caption{Runtime of the precise online algorithm, the interval approximation, and the baseline Lola specification.}
    \label{fig:eval_runtime_interpreter}
\end{subfigure}
\hfill
\begin{subfigure}{0.49\linewidth}
    \includegraphics[width=\textwidth]{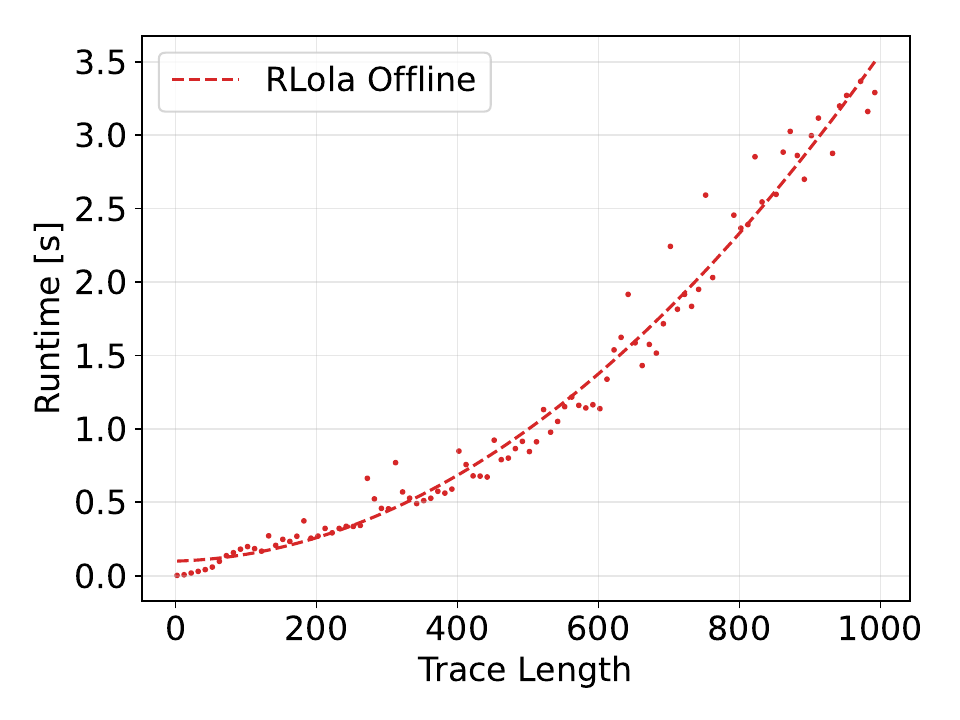}
    \caption{Runtime of the SMT-Solver with growing trace length.\\$\ $}
    \label{fig:runtime_smt}
\end{subfigure}
\caption{Running time of the online and offline algorithms.}
\label{fig:eval_runtime}
\end{figure*}

\Cref{fig:eval_runtime} illustrates the performance of the algorithms across varying trace lengths.
For offline analysis, we use Z3~\cite{DBLP:conf/tacas/MouraB08} and report its runtime via its statistics interface.
Plain Lola execution, which does not account for error margins, is the fastest.
The interval approximation ranks second, as it calculates upper and lower bounds for each stream in the specification, effectively doubling the number of computations.
The precise online algorithm requires further computations to manage slack variables, introducing additional streams and increasing runtime.

Despite this, the runtime of the online algorithms remains small compared to the SMT solver, as shown in \Cref{fig:runtime_smt} (note the different y-axis scale).
The runtime of the online algorithms grows linearly with trace length, whereas the SMT solver's runtime appears to follow an exponential trend.
Overall, the evaluation clearly highlights the inherent trade-off between precision and runtime.

\begin{figure*}
\centering  
\begin{subfigure}{0.49\linewidth}
    \includegraphics[width=\textwidth]{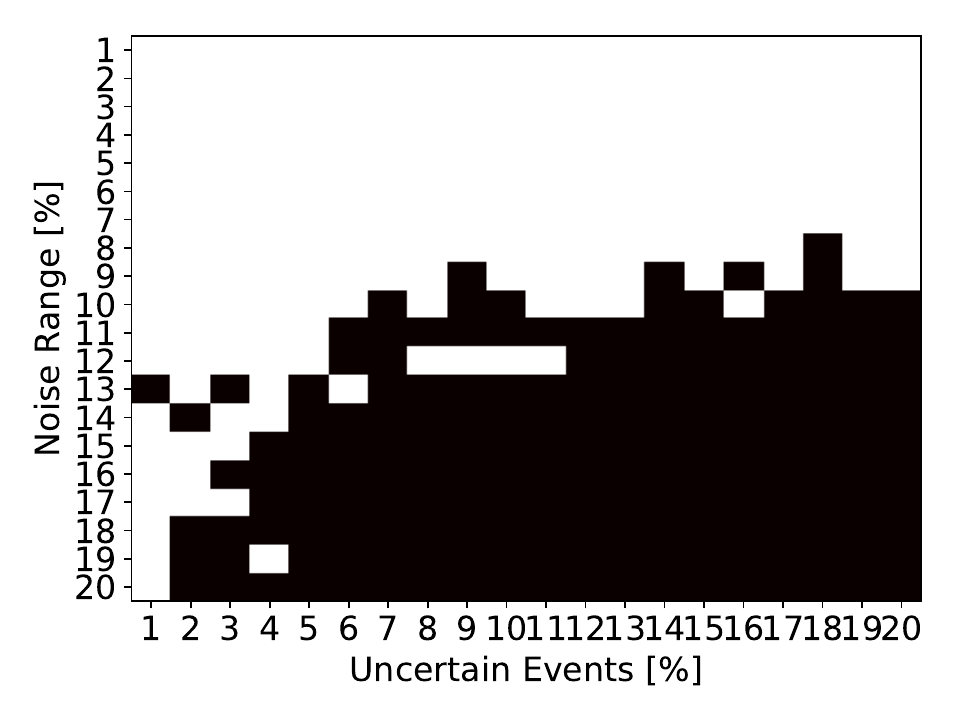}
    \caption{Validity of the trip under varying noise conditions. Black rectangles indicate cases where the monitor does not produce the expected verdict.}
    \label{fig:rde:trip_valid}
\end{subfigure}
\hfill
\begin{subfigure}{0.49\linewidth}
    \includegraphics[width=\textwidth]{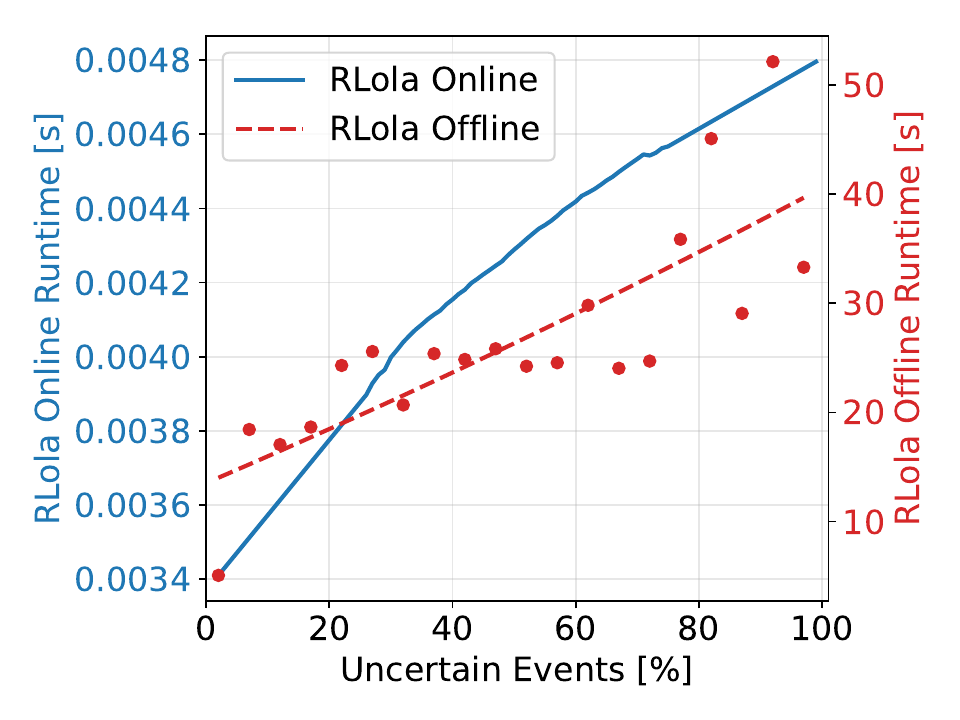}
    \caption{Runtime of the online and offline monitors for different levels of noise.}
    \label{fig:rde:runtime}
\end{subfigure}
\caption{Precision and runtime analysis of the online and offline algorithms on the real driving emission benchmark from~\cite{DBLP:conf/tacas/BiewerFHKSS21}.}
\label{fig:rde}
\end{figure*}

\paragraph{Real Driving Emissions.}
We further evaluate the algorithms on a benchmark from~\cite{DBLP:conf/tacas/BiewerFHKSS21}, where a monitor checks whether a vehicle drive satisfies real driving emission (RDE) requirements.
These requirements include constraints on highway driving, speed ranges, and driving in rural and urban areas.
An excerpt of the specification is shown in \Cref{spec:rde}.

The specification used in our evaluation is derived from the original case study and checks whether the conditions for a valid test drive are met.

\begin{lstlisting}[caption={An excert of the RDE specification.}, label={spec:rde}]
output trip_valid :=
    (90.0 * 60.0) <= t)   && (t <= (120.0 * 60.0) &&
    (0.29 * d) $<_{1.0}$ d_u_l && d_u_m $<_{1.0}$ (0.44 * d) &&
    (0.23 * d) $<_{1.0}$ d_r_l && d_r_m $<_{1.0}$ (0.43 * d) &&
    (0.23 * d) $<_{1.0}$ d_m_l && d_m_m $<_{1.0}$ (0.43 * d) &&
    u_avg_l $>_{1.0}$ 15.0 && u_avg_m $<_{1.0}$ 40.0 &&
    
    (0.06 * t_u_m) <= u_stop_l_pc &&  
    u_stop_m_pc <= (0.30 * t_u_l) && 
    
    v_above_160_t == 0.0 &&
    v_above_145_t <= (0.03 * t_m_l) &&
    
    v_above_100_t >= (5.0 * 60.0) &&
    
    long_stops >= 5.0 &&
    
    d_u_l $>_{1.0}$ (16.0 * 1000.0) &&
    d_r_l $>_{1.0}$ (16.0 * 1000.0) &&
    d_m_l $>_{1.0}$ (16.0 * 1000.0) &&
    altDelta $<_{1.0}$ 100.0
\end{lstlisting}

Following~\cite{DBLP:journals/infsof/HiplerKLMSW26}, \Cref{fig:rde:trip_valid} shows whether the online algorithm produces the expected verdict under varying noise levels.
We extend the specification with an input stream that is set to $0$ for certain events and to the percentage shown on the y-axis for uncertain events.
Uncertain events are randomly distributed to match the percentage on the x-axis.

Black rectangles in~\Cref{fig:rde:trip_valid} indicate that the monitor was unable to reach the expected verdict, i.e., it rejected the trace as invalid.
The figure shows that the algorithm still computes the expected verdict in most low-noise environments.
Further analysis reveals that the main limiting factor is the upper speed bound: the vehicle must not exceed 145 km/h for more than 3\% of highway driving time.
Since the trace already contains many values near 140 km/h, even small noise can violate this constraint.

\Cref{fig:rde:runtime} presents runtime results for both algorithms on the same specification.
The SMT solver Z3~\cite{DBLP:conf/tacas/MouraB08} does not terminate within one hour on the full trace of about 5500 events.
Therefore, we use a prefix of length 200 for runtime evaluation and vary the number of uncertain events.

Preliminary results indicate that noise magnitude has little effect on runtime.
Both algorithms scale approximately linearly with the number of uncertain events, likely due to the increased number of slack variables.
As expected, the offline algorithm is consistently slower than the online algorithm.

\section{Conclusion}\label{sec:conclusion}
We have presented \rlola, a robust extension of the monitoring framework Lola.
In \rlola, the addition of slack variables allows us to track measurement noise induced by inaccurate sensors throughout computations; boolean verdicts explicitly account for the resulting inaccuracies.
We demonstrated that \rlola monitors require, in general, an unbounded amount of memory.
For online monitoring, we addressed this issue with a complete, but approximate, monitoring algorithm based on interval arithmetic and a construction of fully precise constant-memory monitors for a rich fragment of \rlola. 

The presented offline algorithm encodes the specification as an SMT formula, enabling the evaluation of trigger conditions with the benefit of hindsight.
This retrospective analysis allows measurements obtained after a trigger condition is evaluated to refine the precision of the overall analysis.
Compared to the online algorithms, the offline algorithm can identify trigger violations that remain undetected by the online algorithm.
However, the improved precision and versatility come at the expense of an increase in runtime. 

Lastly, we discussed the implementation of the algorithms in the existing RTLola Framework and demonstrated their effectiveness by evaluating the above methods with respect to running time and precision.
\bibliographystyle{spmpsci}
\bibliography{references}

@inproceedings{Lola,
  author       = {Ben D'Angelo and
                  Sriram Sankaranarayanan and
                  C{\'{e}}sar S{\'{a}}nchez and
                  Will Robinson and
                  Bernd Finkbeiner and
                  Henny B. Sipma and
                  Sandeep Mehrotra and
                  Zohar Manna},
  title        = {{LOLA:} Runtime Monitoring of Synchronous Systems},
  booktitle    = {12th International Symposium on Temporal Representation and Reasoning
                  {(TIME} 2005), 23-25 June 2005, Burlington, Vermont, {USA}},
  pages        = {166--174},
  publisher    = {{IEEE} Computer Society},
  year         = {2005},
  doi          = {10.1109/TIME.2005.26},
  bibsource    = {dblp computer science bibliography, https://dblp.org}
}

@inproceedings{TeSSLa,
  author       = {Lukas Convent and
                  Sebastian Hungerecker and
                  Martin Leucker and
                  Torben Scheffel and
                  Malte Schmitz and
                  Daniel Thoma},
  editor       = {Tiago Massoni and
                  Mohammad Reza Mousavi},
  title        = {TeSSLa: Temporal Stream-Based Specification Language},
  booktitle    = {Formal Methods: Foundations and Applications - 21st Brazilian Symposium,
                  {SBMF} 2018, Salvador, Brazil, November 26-30, 2018, Proceedings},
  series       = {Lecture Notes in Computer Science},
  volume       = {11254},
  pages        = {144--162},
  publisher    = {Springer},
  year         = {2018},
  doi          = {10.1007/978-3-030-03044-5\_10},
  bibsource    = {dblp computer science bibliography, https://dblp.org}
}

@inproceedings{Striver,
  author       = {Felipe Gorostiaga and
                  C{\'{e}}sar S{\'{a}}nchez},
  editor       = {Christian Colombo and
                  Martin Leucker},
  title        = {Striver: Stream Runtime Verification for Real-Time Event-Streams},
  booktitle    = {Runtime Verification - 18th International Conference, {RV} 2018, Limassol,
                  Cyprus, November 10-13, 2018, Proceedings},
  series       = {Lecture Notes in Computer Science},
  volume       = {11237},
  pages        = {282--298},
  publisher    = {Springer},
  year         = {2018},
  doi          = {10.1007/978-3-030-03769-7\_16},
  bibsource    = {dblp computer science bibliography, https://dblp.org}
}

@inproceedings{RTLola,
  author       = {Jan Baumeister and
                  Bernd Finkbeiner and
                  Sebastian Schirmer and
                  Maximilian Schwenger and
                  Christoph Torens},
  editor       = {Shuvendu K. Lahiri and
                  Chao Wang},
  title        = {{RTLola} Cleared for Take-Off: Monitoring Autonomous Aircraft},
  booktitle    = {Computer Aided Verification - 32nd International Conference, {CAV}
                  2020, Los Angeles, CA, USA, July 21-24, 2020, Proceedings, Part {II}},
  series       = {Lecture Notes in Computer Science},
  volume       = {12225},
  pages        = {28--39},
  publisher    = {Springer},
  year         = {2020},
  doi          = {10.1007/978-3-030-53291-8\_3},
  bibsource    = {dblp computer science bibliography, https://dblp.org}
}

@InProceedings{cav4,
author="Baumeister, Jan
and Finkbeiner, Bernd
and Kohn, Florian
and L{\"o}hr, Florian
and Manfredi, Guido
and Schirmer, Sebastian
and Torens, Christoph",
editor="Gurfinkel, Arie
and Ganesh, Vijay",
title="Monitoring Unmanned Aircraft: Specification, Integration, and Lessons-Learned",
booktitle="Computer Aided Verification",
doi="10.1007/978-3-031-65630-9\_10",
year="2024",
publisher="Springer Nature Switzerland",
address="Cham",
pages="207--218",
}

@inproceedings{DonzeMaler10,
  author    = {Alexandre Donz{\'{e}} and
               Oded Maler},
  editor    = {Krishnendu Chatterjee and
               Thomas A. Henzinger},
  title     = {Robust Satisfaction of Temporal Logic over Real-Valued Signals},
  booktitle = {8th International
               Conference on Formal Modeling and Analysis of Timed Systems, {FORMATS} 2010, Klosterneuburg, Austria, September 8-10,
               2010. Proceedings},
  series    = {Lecture Notes in Computer Science},
  volume    = {6246},
  pages     = {92--106},
  publisher = {Springer},
  year      = {2010},
  doi       = {10.1007/978-3-642-15297-9\_9},
  bibsource = {dblp computer science bibliography, https://dblp.org}
}

@book{IntervalAnalysis,
  author       = {Ramon E. Moore and
                  R. Baker Kearfott and
                  Michael J. Cloud},
  title        = {Introduction to Interval Analysis},
  publisher    = {{SIAM}},
  year         = {2009},
  doi          = {10.1137/1.9780898717716},
  isbn         = {978-0-89871-669-6},
  bibsource    = {dblp computer science bibliography, https://dblp.org}
}

@article{RobustSTL,
  author       = {Bernd Finkbeiner and
                  Martin Fr{\"{a}}nzle and
                  Florian Kohn and
                  Paul Kr{\"{o}}ger},
  title        = {A Truly Robust Signal Temporal Logic: Monitoring Safety Properties
                  of Interacting Cyber-Physical Systems under Uncertain Observation},
  journal      = {Algorithms},
  volume       = {15},
  number       = {4},
  pages        = {126},
  year         = {2022},
  doi          = {10.3390/A15040126},
  bibsource    = {dblp computer science bibliography, https://dblp.org}
}

@article{AffineArithmetic,
  author       = {Luiz Henrique de Figueiredo and
                  Jorge Stolfi},
  title        = {Affine Arithmetic: Concepts and Applications},
  journal      = {Numer. Algorithms},
  volume       = {37},
  number       = {1-4},
  pages        = {147--158},
  year         = {2004},
  doi          = {10.1023/B:NUMA.0000049462.70970.B6},
  bibsource    = {dblp computer science bibliography, https://dblp.org}
}

@article{DBLP:journals/sttt/KauffmanHF21,
  author       = {Sean Kauffman and
                  Klaus Havelund and
                  Sebastian Fischmeister},
  title        = {What can we monitor over unreliable channels?},
  journal      = {Int. J. Softw. Tools Technol. Transf.},
  volume       = {23},
  number       = {4},
  pages        = {579--600},
  year         = {2021},
  doi          = {10.1007/S10009-021-00625-Z},
  bibsource    = {dblp computer science bibliography, https://dblp.org}
}

@inproceedings{DBLP:conf/rv/LeuckerSS0T19,
  author       = {Martin Leucker and
                  C{\'{e}}sar S{\'{a}}nchez and
                  Torben Scheffel and
                  Malte Schmitz and
                  Daniel Thoma},
  editor       = {Bernd Finkbeiner and
                  Leonardo Mariani},
  title        = {Runtime Verification for Timed Event Streams with Partial Information},
  booktitle    = {Runtime Verification - 19th International Conference, {RV} 2019, Porto,
                  Portugal, October 8-11, 2019, Proceedings},
  series       = {Lecture Notes in Computer Science},
  volume       = {11757},
  pages        = {273--291},
  publisher    = {Springer},
  year         = {2019},
  doi          = {10.1007/978-3-030-32079-9\_16},
  bibsource    = {dblp computer science bibliography, https://dblp.org}
}

@inproceedings{DBLP:conf/tacas/DeckerLT14,
  author       = {Normann Decker and
                  Martin Leucker and
                  Daniel Thoma},
  editor       = {Erika {\'{A}}brah{\'{a}}m and
                  Klaus Havelund},
  title        = {Monitoring Modulo Theories},
  booktitle    = {Tools and Algorithms for the Construction and Analysis of Systems
                  - 20th International Conference, {TACAS} 2014, Held as Part of the
                  European Joint Conferences on Theory and Practice of Software, {ETAPS}
                  2014, Grenoble, France, April 5-13, 2014. Proceedings},
  series       = {Lecture Notes in Computer Science},
  volume       = {8413},
  pages        = {341--356},
  publisher    = {Springer},
  year         = {2014},
  doi          = {10.1007/978-3-642-54862-8\_23},
  bibsource    = {dblp computer science bibliography, https://dblp.org}
}

@inproceedings{DBLP:conf/atva/KallwiesLS22,
  author       = {Hannes Kallwies and
                  Martin Leucker and
                  C{\'{e}}sar S{\'{a}}nchez},
  editor       = {Ahmed Bouajjani and
                  Luk{\'{a}}s Hol{\'{\i}}k and
                  Zhilin Wu},
  title        = {Symbolic Runtime Verification for Monitoring Under Uncertainties and
                  Assumptions},
  booktitle    = {Automated Technology for Verification and Analysis - 20th International
                  Symposium, {ATVA} 2022, Virtual Event, October 25-28, 2022, Proceedings},
  series       = {Lecture Notes in Computer Science},
  volume       = {13505},
  pages        = {117--134},
  publisher    = {Springer},
  year         = {2022},
  doi          = {10.1007/978-3-031-19992-9\_8},
  bibsource    = {dblp computer science bibliography, https://dblp.org}
}

@inproceedings{DBLP:conf/cav/DonzeFM13,
  author       = {Alexandre Donz{\'{e}} and
                  Thomas Ferr{\`{e}}re and
                  Oded Maler},
  editor       = {Natasha Sharygina and
                  Helmut Veith},
  title        = {Efficient Robust Monitoring for {STL}},
  booktitle    = {Computer Aided Verification - 25th International Conference, {CAV}
                  2013, Saint Petersburg, Russia, July 13-19, 2013. Proceedings},
  series       = {Lecture Notes in Computer Science},
  volume       = {8044},
  pages        = {264--279},
  publisher    = {Springer},
  year         = {2013},
  doi          = {10.1007/978-3-642-39799-8\_19},
  bibsource    = {dblp computer science bibliography, https://dblp.org}
}

@inproceedings{DBLP:conf/ictac/FranzleH05,
  author       = {Martin Fr{\"{a}}nzle and
                  Michael R. Hansen},
  editor       = {Dang Van Hung and
                  Martin Wirsing},
  title        = {A Robust Interpretation of Duration Calculus},
  booktitle    = {Theoretical Aspects of Computing - {ICTAC} 2005, Second International
                  Colloquium, Hanoi, Vietnam, October 17-21, 2005, Proceedings},
  series       = {Lecture Notes in Computer Science},
  volume       = {3722},
  pages        = {257--271},
  publisher    = {Springer},
  year         = {2005},
  doi          = {10.1007/11560647\_17},
  bibsource    = {dblp computer science bibliography, https://dblp.org}
}

@Book{ISO:5725,
  author =       "{ISO}",
  title =        "{ISO\slash IEC 5725:2023}: Accuracy (trueness and precision) of measurement methods and results - Part 1: General principles and definitions",
  publisher =    "International Organization for Standardization",
  address =      "Geneva, Switzerland",
  day =          "1",
  month =        "July",
  year =         "2023",
}

@misc{interpreter,
  title = {{RTLola} Interpreter - Rust Crate on crates.io},
  author = {Jan Baumeister and
    Florian Kohn and
    Stefan Oswald and
    Malte Schledjewski and
    Maximilian Schwenger and
    Frederik Scheerer and
    Marvin Stenger and
    Leander Tentrup},
  year = 2025,
  url = {https://crates.io/crates/rtlola-interpreter},
  urldate = {2025-08-30}
}

@misc{frontend,
  title = {{RTLola} Frontend - Rust Crate on crates.io},
  author = {Jan Baumeister and
    Florian Kohn and
    Stefan Oswald and
    Malte Schledjewski and
    Maximilian Schwenger and
    Frederik Scheerer and
    Marvin Stenger and
    Leander Tentrup},
  year = 2025,
  url = {https://crates.io/crates/rtlola-frontend},
  urldate = {2024-05-24}
}

@inproceedings{ViscontiEA:IntervalBasedSTRELMonitoring,
  author    = {Ennio Visconti and
               Ezio Bartocci and
               Michele Loreti and
               Laura Nenzi},
  editor    = {S. Arun{-}Kumar and
               Dominique M{\'{e}}ry and
               Indranil Saha and
               Lijun Zhang},
  title     = {Online monitoring of spatio-temporal properties for imprecise signals},
  booktitle = {19th {ACM-IEEE} International Conference on Formal
               Methods and Models for System Design, Virtual Event, China, November
               20 - 22, 2021},
  publisher = {{ACM}},
  year      = {2021},
  url       = {https://doi.org/10.1145/3487212.3487344},
  doi       = {10.1145/3487212.3487344},
pages = {78–88},
numpages = {11}
}

@article{DBLP:journals/tosem/BauerLS11,
  author       = {Andreas Bauer and
                  Martin Leucker and
                  Christian Schallhart},
  title        = {Runtime Verification for {LTL} and {TLTL}},
  journal      = {{ACM} Trans. Softw. Eng. Methodol.},
  volume       = {20},
  number       = {4},
  pages        = {14:1--14:64},
  year         = {2011},
  doi          = {10.1145/2000799.2000800},
  bibsource    = {dblp computer science bibliography, https://dblp.org}
}

@inproceedings{DBLP:conf/formats/MalerN04,
  author       = {Oded Maler and
                  Dejan Nickovic},
  editor       = {Yassine Lakhnech and
                  Sergio Yovine},
  title        = {Monitoring Temporal Properties of Continuous Signals},
  booktitle    = {Formal Techniques, Modelling and Analysis of Timed and Fault-Tolerant
                  Systems, Joint International Conferences on Formal Modelling and Analysis
                  of Timed Systems, {FORMATS} 2004 and Formal Techniques in Real-Time
                  and Fault-Tolerant Systems, {FTRTFT} 2004, Grenoble, France, September
                  22-24, 2004, Proceedings},
  series       = {Lecture Notes in Computer Science},
  volume       = {3253},
  pages        = {152--166},
  publisher    = {Springer},
  year         = {2004},
  url          = {https://doi.org/10.1007/978-3-540-30206-3\_12},
  doi          = {10.1007/978-3-540-30206-3\_12},
  bibsource    = {dblp computer science bibliography, https://dblp.org}
}

@inproceedings{DBLP:journals/entcs/ThatiR05,
  author       = {Prasanna Thati and
                  Grigore Rosu},
  editor       = {Klaus Havelund and
                  Grigore Rosu},
  title        = {Monitoring Algorithms for Metric Temporal Logic Specifications},
  booktitle    = {Proceedings of the Fourth Workshop on Runtime Verification, RV@ETAPS
                  2004, Barcelona, Spain, April 3, 2004},
  series       = {Electronic Notes in Theoretical Computer Science},
  volume       = {113},
  pages        = {145--162},
  publisher    = {Elsevier},
  year         = {2004},
  doi          = {10.1016/J.ENTCS.2004.01.029},
  bibsource    = {dblp computer science bibliography, https://dblp.org}
}

@inproceedings{barrett2010smt,
  title={The smt-lib standard: Version 2.0},
  author={Barrett, Clark and Stump, Aaron and Tinelli, Cesare and others},
  booktitle={Proceedings of the 8th international workshop on satisfiability modulo theories (Edinburgh, UK)},
  volume={13},
  pages={14},
  year={2010}
}

@incollection{DBLP:reference/mc/BarrettT18,
  author       = {Clark W. Barrett and
                  Cesare Tinelli},
  editor       = {Edmund M. Clarke and
                  Thomas A. Henzinger and
                  Helmut Veith and
                  Roderick Bloem},
  title        = {Satisfiability Modulo Theories},
  booktitle    = {Handbook of Model Checking},
  pages        = {305--343},
  publisher    = {Springer},
  year         = {2018},
  doi          = {10.1007/978-3-319-10575-8\_11},
  bibsource    = {dblp computer science bibliography, https://dblp.org}
}

@inproceedings{DBLP:conf/rv/FinkbeinerFKK24,
  author       = {Bernd Finkbeiner and
                  Martin Fr{\"{a}}nzle and
                  Florian Kohn and
                  Paul Kr{\"{o}}ger},
  editor       = {Erika {\'{A}}brah{\'{a}}m and
                  Houssam Abbas},
  title        = {Stream-Based Monitoring Under Measurement Noise},
  booktitle    = {Runtime Verification - 24th International Conference, {RV} 2024, Istanbul,
                  Turkey, October 15-17, 2024, Proceedings},
  series       = {Lecture Notes in Computer Science},
  volume       = {15191},
  pages        = {22--39},
  publisher    = {Springer},
  year         = {2024},
  doi          = {10.1007/978-3-031-74234-7\_2},
  bibsource    = {dblp computer science bibliography, https://dblp.org}
}

@inproceedings{DBLP:conf/fm/BaumeisterFKS24,
  author       = {Jan Baumeister and
                  Bernd Finkbeiner and
                  Florian Kohn and
                  Frederik Scheerer},
  editor       = {Andr{\'{e}} Platzer and
                  Kristin Yvonne Rozier and
                  Matteo Pradella and
                  Matteo Rossi},
  title        = {A Tutorial on Stream-Based Monitoring},
  booktitle    = {Formal Methods - 26th International Symposium, {FM} 2024, Milan, Italy,
                  September 9-13, 2024, Proceedings, Part {II}},
  series       = {Lecture Notes in Computer Science},
  volume       = {14934},
  pages        = {624--648},
  publisher    = {Springer},
  year         = {2024},
  doi          = {10.1007/978-3-031-71177-0\_33},
  bibsource    = {dblp computer science bibliography, https://dblp.org}
}

@inproceedings{DBLP:conf/tacas/MouraB08,
  author       = {Leonardo Mendon{\c{c}}a de Moura and
                  Nikolaj S. Bj{\o}rner},
  editor       = {C. R. Ramakrishnan and
                  Jakob Rehof},
  title        = {{Z3:} An Efficient {SMT} Solver},
  booktitle    = {Tools and Algorithms for the Construction and Analysis of Systems,
                  14th International Conference, {TACAS} 2008, Held as Part of the Joint
                  European Conferences on Theory and Practice of Software, {ETAPS} 2008,
                  Budapest, Hungary, March 29-April 6, 2008. Proceedings},
  series       = {Lecture Notes in Computer Science},
  volume       = {4963},
  pages        = {337--340},
  publisher    = {Springer},
  year         = {2008},
  doi          = {10.1007/978-3-540-78800-3\_24},
  bibsource    = {dblp computer science bibliography, https://dblp.org}
}

@inproceedings{DBLP:conf/tacas/BiewerFHKSS21,
  author       = {Sebastian Biewer and
                  Bernd Finkbeiner and
                  Holger Hermanns and
                  Maximilian A. K{\"{o}}hl and
                  Yannik Schnitzer and
                  Maximilian Schwenger},
  editor       = {Jan Friso Groote and
                  Kim Guldstrand Larsen},
  title        = {{RTLola} on Board: Testing Real Driving Emissions on your Phone},
  booktitle    = {Tools and Algorithms for the Construction and Analysis of Systems
                  - 27th International Conference, {TACAS} 2021, Held as Part of the
                  European Joint Conferences on Theory and Practice of Software, {ETAPS}
                  2021, Luxembourg City, Luxembourg, March 27 - April 1, 2021, Proceedings,
                  Part {II}},
  series       = {Lecture Notes in Computer Science},
  pages        = {365--372},
  publisher    = {Springer},
  year         = {2021},
  url          = {https://doi.org/10.1007/978-3-030-72013-1\_20},
  doi          = {10.1007/978-3-030-72013-1\_20},
  bibsource    = {dblp computer science bibliography, https://dblp.org}
}

@article{DBLP:journals/infsof/HiplerKLMSW26,
  author       = {Raik Hipler and
                  Hannes Kallwies and
                  Martin Leucker and
                  Marco Montali and
                  C{\'{e}}sar S{\'{a}}nchez and
                  Sarah Winkler},
  title        = {Symbolic runtime verification for monitoring under uncertainties and
                  assumptions},
  journal      = {Inf. Softw. Technol.},
  volume       = {191},
  pages        = {108004},
  year         = {2026},
  doi          = {10.1016/J.INFSOF.2025.108004},
  bibsource    = {dblp computer science bibliography, https://dblp.org}
}

\end{document}